\documentclass[aps,prl,10pt,twocolumn,superscriptaddress,amsmath,amssymb,floatfix]{revtex4-2}

\usepackage{graphicx}
\graphicspath{{figures/}}
\usepackage{dcolumn}
\usepackage{bm}
\usepackage{multirow}
\usepackage{color}
\usepackage{braket}
\usepackage[utf8]{inputenc}
\usepackage{siunitx}
\usepackage{hyperref}
\usepackage{xcolor}
\usepackage{tikz}

\newcommand{\ketbra}[2]{\ket{#1}\!\bra{#2}}
\newcommand{\Ybion}{\ensuremath{^{171}\mathrm{Yb}^+}~}
\newcommand{\Yb}{\ensuremath{^{171}\mathrm{Yb}}~}

\begin{document}

\title{Deterministic atom-shuttle interconnects via ultrafast atom--ion entangling gate}

\author{Mu Qiao}
\email{mu.q.phys@gmail.com}
\affiliation{Universit\'{e} Paris-Saclay, Institut d'Optique Graduate School, CNRS, Laboratoire Charles Fabry, 91127 Palaiseau Cedex, France}

\date{\today}

\begin{abstract}
Neutral-atom arrays and trapped-ion crystals offer complementary strengths for fault-tolerant quantum computing but lack a fast way to deterministically interact. Here we propose a controlled-$Z$ gate generated by the charge--induced-dipole ($C_4$) force between a Rydberg-excited atom and a trapped ion, balanced by a spin-dependent optical Magnus force on the ion that closes phase-space trajectories within a few microseconds. Toggling the Rydberg state extends the scheme to multi-ion crystals at negligible overhead. The resulting ${\sim}5\,$kHz atom shuttle accelerates short-distance QCCD links and enables hybrid qLDPC memories in which atom logical qubits are written onto an ion block treated as a passive storage zone. We perform circuit-level Monte Carlo simulations and find that the hybrid architecture supports orders of magnitude more operations than atom-only or ion-only architectures at fixed code distance and logical error rate.
\end{abstract}

\maketitle

Neutral-atom arrays and trapped-ion crystals are leading platforms for quantum information processing. Atoms in optical tweezers offer fast, parallel Rydberg-mediated gates~\cite{Scholl2023,Ma2023}; ions provide coherence times exceeding ten hours~\cite{Wang2021,Pi2026}, reliable long-term storage, and the highest single- and two-qubit gate fidelities~\cite{Smith2025, Hughes2025}. The ultimate quantum information processor may combine these platforms, using atoms as fast processors or flying interconnects~\cite{Kotochigova2025} and ions as robust quantum memories, but realizing this hybrid architecture requires a fast, deterministic atom--ion entangling gate. Existing proposals rely on slow collisional gates (${\sim}\!300\,\mu$s)~\cite{Kotochigova2025} or millisecond-scale Rydberg-dressed interactions~\cite{Secker2016}; reaching the kHz rates required for a practical interface remains an open challenge~\cite{Monroe2014}.

\begin{figure*}[t]
    \centering
    \includegraphics[width=\textwidth]{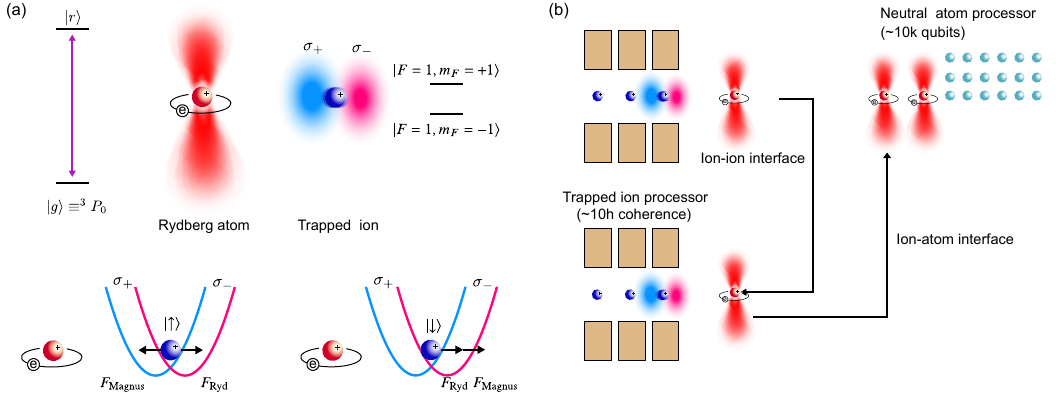}
    \caption{\label{fig:scheme}
    \textbf{Atom--ion interface and hybrid quantum architecture.} (a) Schematic of the entangling gate. The neutral \Yb atom uses a metastable clock state $\ket{g} \equiv {}^3P_0$ and a Rydberg state $\ket{r}$ as the qubit. The \Ybion ion is encoded in stretched Zeeman states ($\ket{F=1, m_F=\pm 1}$) to use spin-dependent optical Magnus forces. (b) Hybrid compute-store quantum architecture. Trapped-ion modules (left) serve as robust quantum memories with demonstrated $\sim\!10\,$h coherence times~\cite{Pi2026}, using standard ion--ion quantum gates for local operations. Neutral-atom arrays (right) provide parallel processing at the thousand-qubit scale~\cite{Manetsch2024}. The proposed hybrid gate gives a deterministic ion--atom interface, letting atoms act as flying interconnects between separated ion modules or transfer processed data into long-term storage.}
\end{figure*}

In this Letter, we propose using the charge--induced-dipole interaction to generate state-dependent forces that rapidly entangle the two particles via geometric phase. Combining this force with a local optical Magnus force on the ion closes phase-space trajectories and realizes a controlled-$Z$ (CZ) gate in a few microseconds [Fig.~\ref{fig:scheme}(a)]. The electric field of an ion at distance $d$ polarizes a nearby atom strongly when the atom is in a $nS$ Rydberg state producing a charge--induced-dipole potential $V = -C_4/d^4$~\cite{Cote2000,Secker2016,Tomza2019,Ewald2019,Engel2018}. For a Rydberg state the resulting force is attractive, conditioned on the atom's qubit state~\cite{Emperauger2025}: when the atom is in $\ket{r}$ the ion is pulled regardless of its spin, leaving $\ket{g}$ unperturbed. To make the interaction also depend on the spin state of the ion we add a static, spin-dependent force on the ion via the optical Magnus effect of a tightly focused beam with a transverse polarization gradient~\cite{Mazzanti2023,Cui2025}. With the ion encoded in magnetically sensitive \emph{stretched} states ($\ket{\!\uparrow},\ket{\!\downarrow}$), the Magnus force pushes it in opposite directions depending on its spin.

We trap the Rydberg atom and tune its secular frequency to match the ion's axial mode $\omega$. After linearizing the $C_4$ potential in the small parameter $\ell_{\text{ion}}/d\!\ll\!1$ (with $\ell_{\text{ion}}\!=\!\sqrt{\hbar/2 m_{\text{ion}}\omega}$ the ion motional zero-point length) and matching the Magnus amplitude to $\omega_g\!\equiv\!4C_4\ell_{\text{ion}}/(\hbar d^5)$, the interaction reduces to a state-dependent constant force on each oscillator (see SM),
\begin{equation}\label{eq:H}
\begin{aligned}
    H = \hbar\omega_g(&a_{\text{ion}}^\dagger{+}a_{\text{ion}})\bigl(\ketbra{r}{r}_{\text{atom}} - \hat{\sigma}_{\text{ion}}^{(z)}\bigr)\\ &- \hbar\omega_g(a_{\text{atom}}^\dagger{+}a_{\text{atom}})\ketbra{r}{r}_{\text{atom}},
\end{aligned}
\end{equation}
where $a_{\text{ion}},a_{\text{ion}}^\dagger$ ($a_{\text{atom}},a_{\text{atom}}^\dagger$) are the ion (atom) motional ladder operators, $\ketbra{r}{r}_{\text{atom}}$ projects onto the Rydberg state of the atom, and $\hat{\sigma}_{\text{ion}}^{(z)}\!\equiv\!\ketbra{\!\uparrow}{\!\uparrow}-\ketbra{\!\downarrow}{\!\downarrow}$ acts on the ion stretched-Zeeman qubit. A constant force on a harmonic oscillator drives a closed circular trajectory in phase space that returns to the origin after one trap period $T{=}2\pi/\omega$, disentangling motion from spin.

\begin{figure*}[t]
    \centering
    \includegraphics[width=\textwidth]{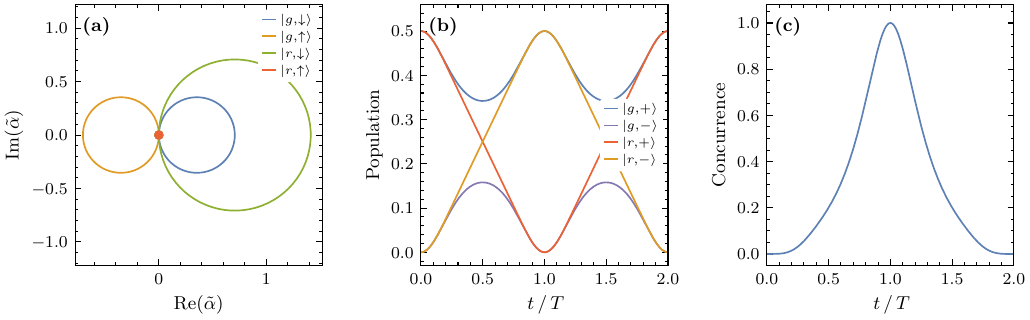}
    \caption{\label{fig:sim}
    \textbf{Single-ion CZ gate.} (a) Phase-space trajectories (rotating frame) of the ion motional mode for each logical state. The $\ket{r,\!\uparrow}$ state remains at the origin (forces cancel), while $\ket{r,\!\downarrow}$ traces the largest loop. (b) Populations in the $\ket{g/r}\!\otimes\!\ket{\pm}$ basis, where $\ket{\pm}=(\ket{\!\downarrow}\pm\ket{\!\uparrow})/\sqrt{2}$. At $t=T$, the conditional phase converts $\ket{r,+}\to\ket{r,-}$, signaling maximal entanglement. (c) Concurrence of the atom--ion spin state, reaching unity at $t=T$ and returning to zero at $t=2T$.}
\end{figure*}

Engineering the Magnus amplitude to match the $C_4$ force, the four logical branches evolve under distinct displacements. With the atom in $\ket{g}$, only the Magnus force acts and the ion is displaced oppositely for $\ket{\!\uparrow}$ and $\ket{\!\downarrow}$; with the atom in $\ket{r}$, the $C_4$ pull cancels the Magnus push on $\ket{\!\uparrow}$ (ion stationary) and reinforces it on $\ket{\!\downarrow}$ (displacement doubled). Each branch traces a distinct closed loop [Fig.~\ref{fig:sim}(a)] and accumulates a geometric phase set by the enclosed area, giving a conditional CZ phase $|\Phi_\mathrm{CZ}| = 8\pi(\omega_g/\omega)^2 = \pi$ at $\omega_g = \omega/(2\sqrt{2})$ [Fig.~\ref{fig:sim}(b,c)]. Residual single-qubit phases are deterministic and removed by local $Z$ rotations.

The gate is completed in a single trap period. Geometric phase gates are intrinsically robust to thermal phonon populations at the linear-Hamiltonian level~\cite{Leibfried2003,Bowers2025}. The anharmonic correction to the $C_4$ potential introduces a weak dependence on the mean ion phonon occupation $\bar n$ that is suppressed to $\sim\!10^{-3}$ by standard sub-Doppler EIT or pulsed-Raman cooling to $\bar n\!\lesssim\!1$ (see Supplemental Material).

For a \Yb atom and \Ybion ion, using the $|6s60p\,{}^3P_2, F\!=\!3/2, m_F\!=\!+3/2\rangle$ Rydberg state, the target conditional phase is achieved at a distance of $d_\mathrm{CZ}\!\approx\!\SI{12}{\micro\meter}$. The primary limit on fidelity is the finite lifetime of the Rydberg state. In a cryogenic apparatus, the radiative-limited lifetime at $n=60$ is $\tau_r\!\approx\!100\text{--}200\,\mu$s~\cite{Beterov2009}; taking the conservative end gives a process-averaged decay error of ${\sim}\!2.4\%$, leaving $\mathcal{F}\!\approx\!97\%$ after technical errors. Moving to $n\!=\!80$ extends $\tau_r$ by $\sim\!2.4\times$ and brings the fidelity above $99\%$. At room temperature, blackbody-induced transitions reduce the effective lifetime by ${\sim}3\text{--}4\times$~\cite{Jenkins2022,Ma2023}, increasing the decay error accordingly; these numbers are tabulated explicitly in the Supplemental Material. Cryogenic trapped-ion setups are standard in the field~\cite{Brownnutt2015,Brandl2016}, and the narrow-line tweezer cooling needed for \Yb (${}^3P_0\!-\!{}^3S_1$, $770\,$nm) is fully compatible with a $4\,$K bath.

Rydberg decay predominantly results in loss of the atom from the tweezer (or autoionization). This provides built-in erasure conversion~\cite{Wu2022,Scholl2023}: decay events are heralded by checking the atom population, converting otherwise unlocated Pauli errors into erasures.

Trapped-ion quantum processors typically operate on 1D or 2D crystals. However, applying a force to one ion in a crystal excites a dense spectrum of collective normal modes (spectator modes). If these modes do not complete full phase-space loops by the end of the gate, residual spin-motion entanglement destroys the gate fidelity. 

In standard M{\o}lmer--S{\o}rensen gates, spectator modes are decoupled by amplitude or frequency modulation~\cite{Leung2018}. We introduce an analogous technique for the atom--ion gate: \emph{Rydberg-state toggling}. During the $5\,\mu$s gate, nanosecond microwave $\pi$-pulses on the $\sim\!5$~GHz $|{}^3P_2, F\!=\!3/2\rangle\!\leftrightarrow\!|{}^3D_2, F\!=\!3/2\rangle$ transition drive the atom between an attractive Rydberg state ($6s60p\,{}^3P_2$, $C_4{>}0$) and a repulsive one ($6s60d\,{}^3D_2$, $C_4{<}0$); both polarizabilities are computed from the Peper--Kuroda multichannel quantum-defect model~\cite{Peper2025,Kuroda2025} using \texttt{PairInteraction}~\cite{Mogerle2025} and are anchored by their mutual dipole coupling to give a matched $|C_4|$ ratio of $0.84$ in $^{171}$Yb across the entire $d\!\in\![5,50]\,\mu$m range (Fig.~\ref{fig:alpha_vs_d}, Sec.~S6). Each toggle flips the sign of the $C_4$ force, and an optimized schedule folds the phase-space trajectories of all collective modes so they simultaneously close at $t{=}T$ (Fig.~\ref{fig:crystalS}, SM). Numerical simulations confirm that this maintains ${\geq}\,97.4\%$ fidelity for crystals up to 10 ions with negligible time overhead ($<\!350\,$ns).

\begin{figure}[t]
    \centering
    \includegraphics[width=\columnwidth]{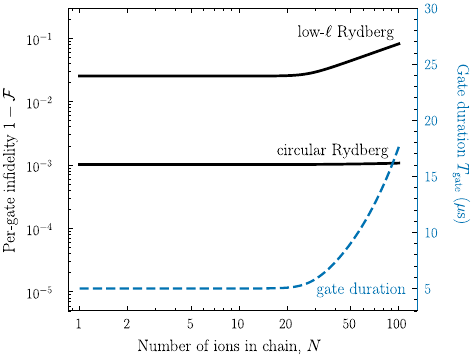}
    \caption{\label{fig:Nscan}
    \textbf{Per-gate infidelity vs.\ ion-chain size $N$.} Orange: low-$\ell$ ${}^3P_2\!\leftrightarrow\!{}^3D_2$ toggle pair. Green: circular Rydberg. Both at $4\,$K, EIT-cooled $\bar n\!=\!1$, standard technical budget (Table~\ref{tab:errors}). Blue dashed (right axis): required gate duration $T_\mathrm{gate}(N)$, set by the toggle-schedule feasibility constraint $\Delta t_\mathrm{min}\!\geq\!\tau_\mathrm{pulse}\!=\!20\,$ns. For $N\!\lesssim\!25$ the schedule fits within one trap period; beyond, $T_\mathrm{gate}$ grows linearly with $N$ and the decay contribution $\tfrac{1}{2}(1\!-\!e^{-T_\mathrm{gate}/\tau_r})$ follows. The circular upgrade suppresses decay by three orders of magnitude, so the low-$\ell$ degradation with $N$ disappears.}
\end{figure}

Figure~\ref{fig:Nscan} summarizes the per-gate infidelity versus ion-chain length $N$ at $4\,$K, for both Rydberg choices, alongside the required gate duration $T_\mathrm{gate}(N)$. At small $N$ the gate completes in one trap period and the infidelity is $N$-independent, saturating at the per-ion Rydberg-decay floor. Above a crossover $N_\star\!\approx\!25$ set by the microwave $\pi$-pulse width ($\tau_\mathrm{pulse}\!\approx\!20\,$ns), the ${\sim}\!2N$ toggle segments of the mode-closure schedule (see SM) no longer fit inside one trap period and $T_\mathrm{gate}$ grows linearly with $N$. For the low-$\ell$ $^3P_2\!\leftrightarrow\!^3D_2$ pair ($\tau_r\!\approx\!100\,\mu$s) this pushes the decay error from $2.5\%$ at $N{=}1$ to ${\sim}\!7\%$ at $N{=}100$ and dominates the $N$-dependence. Circular Rydberg states ($\tau_r\!\approx\!120\,$ms, see SM) suppress the decay floor by three orders of magnitude, so the same ${\sim}2.5{\times}$ increase in gate duration raises decay only from $2.1{\times}10^{-5}$ to $5{\times}10^{-5}$, negligible against the ${\sim}\!10^{-3}$ technical budget. The circular point therefore gives $\mathcal{F}\!\approx\!99.9\%$ across the full range $N\!\in\![1,100]$, well within the fault-tolerant logical-transfer regime~\cite{Gu2025BellQLDPC}. Unlike standard M{\o}lmer--S{\o}rensen gates, where pair-wise constraints scale as $N(N{-}1)/2$, our protocol addresses a single target ion: the only $N$-dependence enters through $T_\mathrm{gate}$.

The microsecond gate time makes neutral atoms usable as deterministic ``flying qubits'' between physically separated trapped-ion modules. In this protocol, an atom is trapped in a movable optical tweezer. It performs a CZ gate with an ion in Module A, is physically shuttled to Module B at speeds of $v \approx \SI{0.5}{\micro\meter/\micro\second}$~\cite{Barredo2016}, and performs a second CZ gate with an ion there. Measuring the atom in the $\{|g\rangle, |r\rangle\}$ basis heralds a maximally entangled Bell state between the two distant ions via entanglement swapping.

\begin{figure}[t]
    \centering
    \includegraphics[width=\columnwidth]{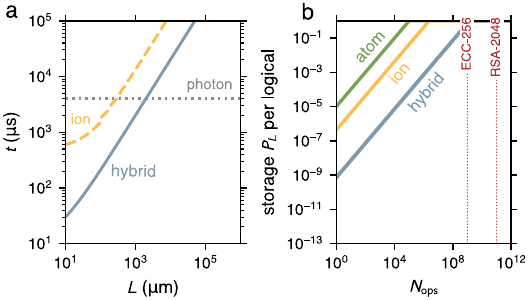}
    \caption{\label{fig:hybrid}
    \textbf{Hybrid architecture advantages.} (a) Interconnect time vs.\ separation $L$ for atom shuttle, QCCD ion shuttle, and heralded photonic link~\cite{OReilly2024}; atom shuttle dominates out to ${\sim}\!2\,$mm. (b) Storage logical error per algorithm operation, $P_L^\mathrm{store}/N_\mathrm{ops}$, for BB$[[72,12,6]]$ at $99.9\%$ atom CZ. Pure-atom and pure-ion architectures pay a constant per-op cost set by their continuous syndrome extraction. Hybrid memory error is the one-shot transfer cost $2p_T$, amortized over $N_\mathrm{ops}$, so the per-op cost decreases with algorithm length and overtakes pure ion at $N_\mathrm{ops}\!\sim\!10^3$. Vertical lines: ECC-256 and RSA-2048 op budgets~\cite{Cain2026Shor}.}
\end{figure}

If atoms are pre-cooled and extracted from a reservoir in their long-lived ground state, the cycle time is set by one-way tweezer transport (${\sim}\!\SI{200}{\micro s}$ over an $L\!=\!\SI{100}{\micro\meter}$ gap) plus the $5\,\mu$s gate, giving an entanglement rate of ${\sim}\!5\,$kHz. As shown in Fig.~\ref{fig:hybrid}(a), atom shuttling dominates every other deterministic interconnect up to ${\sim}\!2\,$mm separation, beyond which heralded photonic links overtake it. The resulting rate is $20\times$ that of state-of-the-art heralded photonic links~\cite{OReilly2024}, and the scheme avoids the heating and time overhead of splitting and shuttling ion chains in QCCD architectures~\cite{Pino2021}. Beyond modular interconnects, the same gate also supports a single-processor heterogeneous architecture, where atom arrays execute parallel local computation and the results are deterministically transferred to ion crystals for long-term storage.

While the gate requires the ion to be temporarily encoded in magnetically sensitive stretched states ($m_F\!=\!\pm 1$), a single pulse maps it back to the clock transition ($\ket{F{=}0,m_F{=}0}\!\leftrightarrow\!\ket{F{=}1,m_F{=}0}$), where the coherence time exceeds ten hours~\cite{Wang2021,Pi2026}. This enables a hybrid memory whose logical error is bounded by the write/read transfer rather than the storage duration [Fig.~\ref{fig:hybrid}(b)]. A pure-atom BB qLDPC memory~\cite{Bravyi2024} must run continuous syndrome extraction (SE) throughout $T_\mathrm{store}$, with logical error growing linearly in the number of rounds~\cite{Breuckmann2021,Bravyi2024}. The hybrid scheme instead writes the state onto an ion BB block via the stabilizer-projection protocol of Ref.~\cite{Gu2025BellQLDPC} and reads it back via the inverse, treating the clock-state ion block as a passive storage zone in the zoned-architecture sense of Refs.~\cite{Bluvstein2025,Cain2026Shor} where SE is used only when an operation requires it, and the long $T_2^\mathrm{ion}$ protects the encoded data in between. With our atom--ion gate and a near-term atom--atom Rydberg CZ at $p_\mathrm{aa}{=}1{\times}10^{-3}$, circuit-level Monte Carlo (see SM) gives a write/read logical Bell-pair infidelity $p_T{=}2.67{\times}10^{-4}$ in BB $[[72,12,6]]$ at $6$ atom--ion gates per pair, falling to $p_T{=}2.22{\times}10^{-5}$ at $d{=}12$ ($12$ gates per pair). Independent code-capacity simulation of the ion block under realistic clock-state decoherence ($T_2^\mathrm{ion}\!=\!10\,$h~\cite{Wang2021,Pi2026}) places the simulated passive-storage logical lifetime at $T_\mathrm{store}^\star\!\approx\!360\,$s before the idle contribution exceeds the transfer budget $2p_T$ (see SM). The atom--ion interface is used only at write/read events, so the cross-platform gate budget is independent of $T_\mathrm{store}$ across the relevant range.

Such idle intervals are structural in fault-tolerant workflows~\cite{Cain2026Shor,Liyanage2023}: lattice-surgery logical gates and decoder latency leave each non-participating qubit idle for ${\sim}\!100\,$ms per logical gate; magic-state factories run an order of magnitude below the Clifford rate, idling consumers for seconds per consumed magic state; and the zoned architectures required for cryptographically relevant problems route ${\sim}\!10^5$ logical qubits through only ${\sim}\!10^3$ processors~\cite{Cain2026Shor}, giving each memory qubit an average duty cycle ${\lesssim}\!1\%$ over the $10$ to $100$ day ECC-256/RSA-2048 wall-clock runtime. During these idle periods a pure-atom memory must keep running SE rounds, whereas clock-state ion storage requires none.

By replacing slow energy penalties with fast state-dependent forces, the charge--induced-dipole interaction gives deterministic atom--ion entanglement in microseconds, with experimental requirements within current capabilities~\cite{Brownnutt2015,Brandl2016,Pultinevicius2025}. The circular-Rydberg operating point of Fig.~\ref{fig:Nscan} (see SM) further raises the projected fidelity to ${\geq}\,99.85\%$ at room temperature and ${\geq}\,99.97\%$ in a $4\,$K cryogenic enclosure, by using the Purcell-suppressed lifetime shown in Ref.~\cite{Pultinevicius2025}. Together with the ten-hour ion clock-state coherence, the gate supports both the modular interconnects and the hybrid qLDPC memories analyzed above within a single apparatus.

\bibliography{main}
\bibliographystyle{apsrev4-2}

\subsection*{Acknowledgements}
We thank Ye Wang, Yao Lu, Romain Martin, Bastien Gély, Yuki Torii Chew, Daniel Barredo, Thierry Lahaye, and Antoine Browaeys for fruitful discussions.

%% =========================================================
%%  SUPPLEMENTAL MATERIAL
%% =========================================================
\clearpage
\onecolumngrid
\setcounter{section}{0}
\setcounter{equation}{0}
\setcounter{figure}{0}
\setcounter{table}{0}
\renewcommand{\theequation}{S\arabic{equation}}
\renewcommand{\thefigure}{S\arabic{figure}}
\renewcommand{\thetable}{S\arabic{table}}

\vspace{2em}
\begin{center}
{\large\textbf{Supplemental Material}}\\[0.5em]
{\large Deterministic atom-shuttle interconnects via ultrafast atom--ion entangling gate}
\end{center}
\vspace{1em}

This Supplemental Material provides the derivations and technical details supporting the physical mechanisms described in the main text. We detail the Hamiltonian derivation, the geometric phase condition, the calculation of the $C_4$ coefficient, the complete error budget, the optical Magnus force and the laser-power requirement to implement it, the multi-ion Rydberg-toggling protocol, an MQDT-based selection of the microwave-coupled $|^3P_2, F\!=\!3/2\rangle\!\leftrightarrow\!|^3D_2, F\!=\!3/2\rangle$ toggle pair, thermal robustness, and the circuit-level validation of the qLDPC logical-transfer protocol used to quantify the hybrid-architecture advantage in Fig.~\ref{fig:hybrid}(b).

\subsection*{S1. Hamiltonian and Phase-Space Dynamics}

We consider an atom and an ion separated by a baseline distance $d$ along the trap axis $x$. The $C_4$ interaction potential is $V = -C_4/|d + x_{\text{ion}} - x_{\text{atom}}|^4$, where $x_{\text{ion}}$ and $x_{\text{atom}}$ are the small excursions of the ion and atom from their equilibrium positions. Because the motional zero-point spreads are small compared to the separation ($\ell/d \sim 10^{-3}$), we linearize the potential to first order in $x/d$, yielding the interaction Hamiltonian:
\begin{equation}\label{eq:HC4}
    H_{C_4} = \frac{4C_4}{d^5}\hat{x}_{\text{ion}}\ketbra{r}{r}_{\text{atom}} - \frac{4C_4}{d^5}\hat{x}_{\text{atom}}\ketbra{r}{r}_{\text{atom}}\,,
\end{equation}
where $\hat{x}_{\text{ion}} = \ell_{\text{ion}}(a_{\text{ion}}^\dagger{+}a_{\text{ion}})$ and $\hat{x}_{\text{atom}} = \ell_{\text{atom}}(a_{\text{atom}}^\dagger{+}a_{\text{atom}})$ are the position operators, $\ell_{\text{ion}}\!=\!\sqrt{\hbar/2 m_{\text{ion}}\omega}$ and $\ell_{\text{atom}}\!=\!\sqrt{\hbar/2 m_{\text{atom}}\omega}$ the corresponding motional zero-point lengths, and the relative minus sign is a direct consequence of Newton's third law (attractive force).

Simultaneously, a tightly focused beam with a transverse polarization gradient generates an optical Magnus effect on the ion~\cite{Mazzanti2023,Cui2025}. This creates a constant force proportional to the ion spin projection $\hat{\sigma}_{\text{ion}}^{(z)}$:
\begin{equation}\label{eq:HMag}
    H_\mathrm{Mag} = -\hbar\Omega_\mathrm{Mag}(a_{\text{ion}}^\dagger{+}a_{\text{ion}})\hat{\sigma}_{\text{ion}}^{(z)}\,,
\end{equation}
where $\Omega_\mathrm{Mag} = g\ell_{\text{ion}}/\hbar$ with $g = 4\tilde{U}_0\bar{\lambda}/w_0^2$. By tuning the laser intensity such that the Magnus amplitude matches the $C_4$ amplitude, $\Omega_\mathrm{Mag} = \omega_g \equiv 4C_4\ell_{\text{ion}}/(\hbar d^5)$, the combined Hamiltonian becomes:
\begin{equation}\label{eq:Heff}
    H = \hbar\omega_g(a_{\text{ion}}^\dagger{+}a_{\text{ion}})\bigl(\ketbra{r}{r}_{\text{atom}} - \hat{\sigma}_{\text{ion}}^{(z)}\bigr) - \hbar\omega_g(a_{\text{atom}}^\dagger{+}a_{\text{atom}})\ketbra{r}{r}_{\text{atom}}\,.
\end{equation}

When acting on a specific logical branch $\ket{k,s}$ (where $k\in\{g,r\}$ and $s\in\{\uparrow,\downarrow\}$), this Hamiltonian reduces to a constant displacement force on the ion mode:
\begin{equation}
    H_{k,s} = \hbar f_{k,s}\, \omega \,(a_{\text{ion}}^\dagger + a_{\text{ion}})\,,
\end{equation}
where $f_{k,s}$ is a dimensionless force amplitude. In the rotating frame at trap frequency $\omega$, a constant force $f$ drives a coherent state $|\alpha(t)\rangle$ along a circular trajectory with radius $|\alpha|_\mathrm{max} = |f|$. This loop perfectly closes after one trap period $T = 2\pi/\omega$. The enclosed area in phase space generates a geometric phase:
\begin{equation}\label{eq:geom}
    \phi_{k,s} = 2\pi f_{k,s}^2\,.
\end{equation}

Evaluating the Hamiltonian~\eqref{eq:Heff} on each branch, the dimensionless force amplitudes are $f_{g\downarrow}=+\omega_g/\omega$, $f_{g\uparrow}=-\omega_g/\omega$, $f_{r\downarrow}=+2\omega_g/\omega$, and $f_{r\uparrow}=0$. The force cancellation for $\ket{r,\!\uparrow}$ directly reflects the matched Magnus and $C_4$ amplitudes. Applying Eq.~\eqref{eq:geom}, the conditional logic phase is:
\begin{equation}
    \Phi_\mathrm{CZ} = \phi_{g\downarrow} + \phi_{r\uparrow} - \phi_{g\uparrow} - \phi_{r\downarrow} = -8\pi(\omega_g/\omega)^2\,.
\end{equation}
Setting $|\Phi_\mathrm{CZ}| = \pi$ gives the required coupling strength $\omega_g = \omega/(2\sqrt{2})$, which dictates the target operating distance $d_\mathrm{CZ} = \left(8\sqrt{2}\,C_4\ell_{\text{ion}} / \hbar\omega\right)^{1/5}$. The residual single-qubit phases are deterministic and trivially removed by standard local $Z$ rotations.

\subsection*{S2. $C_4$ Coefficient Calculation}

The static polarizability of a Rydberg state $|n,l\rangle$, with effective quantum number $\nu^* = n - \delta_l$ (where $\delta_l$ is the angular-momentum-dependent quantum defect), is given by second-order perturbation theory~\cite{Gallagher1994}:
\begin{equation}\label{eq:alpha}
    \alpha_0 = \sum_{n',l'}\frac{2|\langle n,l | e\,\hat r_e | n',l' \rangle|^2}{E_{n'l'} - E_{nl}}\,,
\end{equation}
where $\hat r_e$ is the radial coordinate operator of the Rydberg electron (not to be confused with the qubit-state label $\ket{r}$ of the main text), $E_{nl}$ the unperturbed energy of $\ket{n,l}$, and the sum runs over all dipole-allowed intermediate bound states $\ket{n',l'}$. This scales as $\nu^{*7}$ and sets the interaction strength $C_4 = \alpha_0 e^2/[2(4\pi\epsilon_0)^2]$. For a precise estimate, we use the open-source ARC (Alkali Rydberg Calculator) package~\cite{ARC}. Evaluating the Stark map for $^{87}$Rb at $n=60$ gives $C_4(60S_{1/2}) = 1.23\times10^{-47}$~J\,m$^4$. 

For the \Yb atom, the $C_4$ coefficient of the $6s60s\,{}^3\!S_1$ Rydberg state can be approximately estimated by scaling from the Rb calculation using the Yb quantum-defect structure, yielding $C_4(6s60s\,{}^3\!S_1) \approx 1.1\times10^{-47}$~J\,m$^4$. This Rb-scaling estimate is provided for orientation only: it neglects the Yb-specific perturber structure that strongly modifies polarizabilities of low-$\ell$ Rydberg states near FÃ¶rster crossings. The actual operating states for the toggle protocol are $6s60p\,{}^3P_2,F\!=\!3/2$ (attractive) and $6s60d\,{}^3D_2,F\!=\!3/2$ (repulsive), whose $C_4$ values are computed directly from the Peper--Kuroda multichannel quantum-defect models~\cite{Peper2025,Kuroda2025} in Sec.~S6 (Table~\ref{tab:60_polarizabilities}); both have $|C_4|\!\approx\!2\times 10^{-46}$~J\,m$^4$ with O(factor 3) absolute calibration uncertainty. The exact $C_4$ for a specific experimental setup can additionally be calibrated \emph{in situ} by measuring the Rydberg Stark shift in the static Coulomb field of the trapped ion at various distances~\cite{Ewald2019,Engel2018}.

\subsection*{S3. Operating Parameters and Error Budget}

Table~\ref{tab:params} provides the explicit physical parameters for the $n=60$ operating point, assuming a standard secular trap frequency of $\omega/(2\pi) = 200$~kHz. 

\begin{table}[h]
\caption{\label{tab:params}Gate parameters for $^{171}$\Yb--\Ybion at the $n\!=\!60$ operating point. The toggle pair is $|{}^3P_2, F\!=\!3/2, m_F\!=\!+3/2\rangle \leftrightarrow |{}^3D_2, F\!=\!3/2, m_F\!=\!+3/2\rangle$ (Sec.~S6); $C_4$, $d_\mathrm{CZ}$, and ion field reflect the \texttt{PairInteraction}~\cite{Mogerle2025} central value with O(factor 3) calibration uncertainty for these $^{1,3}\!L_2$-mixed states.}
\begin{ruledtabular}
\begin{tabular}{lc}
Parameter & Value \\
\hline
Trap frequency $\omega_{\text{ion}} = \omega_{\text{atom}}$ & $2\pi\times\SI{200}{kHz}$ \\
Atom--ion distance $d_\mathrm{CZ}$ & $\sim\!\SI{12}{\micro\meter}$ \\
Gate time $T = 2\pi/\omega$ & \SI{5.0}{\micro\second} \\
$|C_4|$ (${}^3P_2$) at $\nu\!=\!59.671$ & $1.89\!\times\!10^{-46}$~J\,m$^4$ \\
$|C_4|$ (${}^3D_2$) at $\nu\!=\!59.835$ & $1.59\!\times\!10^{-46}$~J\,m$^4$ \\
$|C_4|$ ratio (${}^3D_2$ / ${}^3P_2$) & $0.84$ (mutual-coupling anchored) \\
Coupling $\omega_g/(2\pi)$ & \SI{71}{kHz} \\
Max.\ displacement $|\alpha|$ & $1/\sqrt{2}$ quanta \\
Ion field at $d_\mathrm{CZ}$ & $\sim\!\SI{10}{V/m}$ \\
Inglis-Teller limit ($n{=}60$) & $\sim\!\SI{660}{V/m}$ \\
Magnus coupling $\Omega_\mathrm{Mag}/(2\pi)$ & \SI{71}{kHz} \\
Toggle drive frequency (${}^3P_2\!\leftrightarrow\!{}^3D_2$) & $\sim\!\SI{5}{GHz}$ \\
Rydberg lifetime $\tau_r$ (${}^3D_2$/${}^3P_2$, cryogenic, 4\,K) & ${\sim}\!\SI{100}{\micro\second}$ (limited by ${}^3D_2$) \\
\end{tabular}
\end{ruledtabular}
\end{table}

Table~\ref{tab:errors} breaks down the detailed error budget for a single-ion gate. Because only half the computational basis (the $\ket{r}$ states) experiences spontaneous decay, the process-averaged decay infidelity is $\frac{1}{2}(1-e^{-T/\tau_r}) \approx 2.4\%$ at the conservative $\tau_r\!\approx\!100\,\mu$s set by the ${}^3D_2$ side of the toggle pair (${}^3P_2$ is longer-lived at $\tau\!\approx\!1.3\,$ms). Thermal position fluctuations ($\sigma_x \approx \SI{6.8}{nm}$ at \SI{3}{\micro K}) cause shot-to-shot variation in the $C_4$ force, slightly breaking the match with the Magnus force. The AC Stark shift of the Magnus beam on the Rydberg atom is negligible ($<10^{-5}$) because the beam is focused to $w_0\approx 1\,\mu$m on the ion, and its intensity at the atom's location ($d_\mathrm{CZ}\!\approx\!12\,\mu$m) falls off exponentially.

\paragraph*{Rydberg lifetime at different blackbody temperatures.}
The $200\,\mu$s Rydberg lifetime quoted in the main text is the $T\!\to\!0$ \emph{radiative} limit for the $6s60s\,^3\!S_1$ state, scaling as $\tau_r^\mathrm{rad}\!\propto\!(n^*)^3$~\cite{Beterov2009}. In a room-temperature apparatus, blackbody-radiation (BBR) induced transitions to neighboring states shorten the effective lifetime by a factor of 3--4 at $n\!\sim\!60$ and still by ${\sim}\!2\!\times$ at $n\!=\!80$. Table~\ref{tab:tau_vs_T} collects lifetime estimates in the two regimes most relevant for a realistic apparatus: a room-temperature (300\,K) chamber and a cryogenic (4\,K) enclosure. The 300\,K numbers are based on experimentally measured Yb $6sns\,^3\!S_1$ lifetimes from Refs.~\cite{Jenkins2022,Ma2023}; the 4\,K numbers follow from suppressing the BBR component and are consistent with the Beterov semiclassical model~\cite{Beterov2009}. The resulting per-gate decay error $\tfrac{1}{2}(1-e^{-T/\tau_r})$ is given in the last column.

\begin{table}[h]
\caption{\label{tab:tau_vs_T}Rydberg-state lifetime $\tau_r$ for representative low-$\ell$ \Yb Rydberg states ($6sns\,^3\!S_1$ measured experimentally; ${}^3D_2$ taken as comparable within a factor of ${\sim}\!2$, dominated by the same $(n^*)^3$ radiative scaling and the same BBR-coupled neighboring states; ${}^3P_2$ is longer-lived by $\sim\!4\times$ from its restricted E1 decay channel structure but is not the binding constraint), and the resulting per-gate decay infidelity $\tfrac{1}{2}(1-e^{-T/\tau_r})$ at $T_\mathrm{gate}\!=\!5\,\mu$s. The 300\,K values are experimental \Yb $6sns$ tweezer numbers; the 4\,K values are the BBR-suppressed radiative limit. The $n\!=\!80$ operating point gains a factor of $(80/60)^3\!\approx\!2.4$ in radiative lifetime on top of the $(80/60)^2\!\approx\!1.8$ reduction in BBR rate, giving the compounded improvement below. For the ${}^3P_2\!\leftrightarrow\!{}^3D_2$ toggle pair of Sec.~S6 the conservative single-state estimate is $\tau_r({}^3D_2)\!\sim\!100\,\mu$s at 4\,K (limiting; ${}^3P_2$ at $\sim\!1.3\,$ms is not the bottleneck), giving the $2.4\%$ decay error of Table~\ref{tab:errors}; the ${}^3S_1$ values listed below thus serve as the optimistic end of the reasonable lifetime range.}
\begin{ruledtabular}
\begin{tabular}{lccc}
Regime & $\tau_r\,(n{=}60)$ & $\tau_r\,(n{=}80)$ & Decay error \\
\hline
Cryogenic, 4\,K (radiative limit)   & $200\,\mu$s  & $470\,\mu$s  & $1.3\%\,/\,0.5\%$ \\
Room temperature, 300\,K            & $\sim\!60\,\mu$s   & $\sim\!150\,\mu$s  & $4.0\%\,/\,1.6\%$ \\
\end{tabular}
\end{ruledtabular}
\end{table}

Combined with the other error channels (Table~\ref{tab:errors}, Table~\ref{tab:thermal_fidelity}) the gate fidelity at $n\!=\!60$ degrades from $\mathcal{F}\!=\!98.4\%$ (cryogenic, EIT-cooled) to $\mathcal{F}\!\approx\!95.6\%$ at room temperature; at $n\!=\!80$ it degrades from $99.4\%$ to $97.8\%$. Room-temperature operation is therefore possible but meaningfully worse; a cryogenic enclosure, routine in modern Paul-trap apparatuses~\cite{Brownnutt2015,Brandl2016}, recovers the radiative-limit fidelity and is assumed throughout the main text. The $87\%$ heralded-erasure fraction of Rydberg decay~\cite{Wu2022,Scholl2023} applies equally at both temperatures, so even the room-temperature $4\%$ decay error contributes only ${\sim}0.5\%$ unheralded Pauli infidelity at $n\!=\!60$, making even 300\,K operation useful within an erasure-aware qLDPC protocol (Sec.~S8).

\begin{table}[h]
\caption{\label{tab:errors}Detailed error budget for the single-ion gate at the ${}^3P_2\!\leftrightarrow\!{}^3D_2$ toggle-pair operating point ($n\!=\!60$, $T\!=\!\SI{5}{\micro\second}$, EIT-cooled $\bar n\!\approx\!0.3$ corresponding to $T_\mathrm{ion}\!=\!3\,\mu$K). The Rydberg-decay row uses $\tau_r\!\approx\!100\,\mu$s as a conservative estimate set by the ${}^3D_2$ side (${}^3P_2$ is longer-lived); experimental verification of these specific lifetimes at $n\!=\!60$ is part of future work. The anharmonic-$C_4$ contribution at this cooling level is ${\lesssim}\,3{\times}10^{-4}$ (Sec.~S7, Table~\ref{tab:thermal_fidelity}) and is already subsumed in the ``thermal position fluctuation'' row below; the $d_\mathrm{CZ}\!\approx\!12\,\mu$m operating point gives an anharmonicity parameter $\eta\!=\!\ell_{\text{ion}}/d$ consistent with the entries in Table~\ref{tab:thermal_fidelity}.}
\begin{ruledtabular}
\begin{tabular}{lc}
Error source & Infidelity ($1-\mathcal{F}$) \\
\hline
Rydberg decay $[\frac{1}{2}(1-e^{-T/\tau_r})]$, $\tau_r\!=\!100\,\mu$s & $2.4\times10^{-2}$ \\
Thermal position fluctuation (\SI{3}{\micro K}) & $\lesssim 6\times10^{-4}$ \\
Trap frequency mismatch ($\delta\omega/\omega = 0.1\%$) & $1\times10^{-5}$ \\
Ion motional heating ($\dot{n} = 100$/s) & $5\times10^{-4}$ \\
Magnus intensity noise ($\delta\Omega/\Omega = 1\%$) & $2\times10^{-4}$ \\
Photon scattering (Magnus beam) & $<10^{-4}$ \\
Micromotion ($\beta = 0.01$) & $1\times10^{-4}$ \\
AC Stark shift (Magnus beam on Rydberg atom) & $<10^{-5}$ \\
\hline
\textbf{Total Process-Averaged Infidelity} & $\mathbf{\sim 2.6\times10^{-2}}$ \\
\end{tabular}
\end{ruledtabular}
\end{table}

\subsection*{S4. Optical Magnus Force and Laser-Power Requirement}

The spin-dependent force on the ion is generated by a tightly focused Gaussian beam with a transverse polarization gradient~\cite{Mazzanti2023,Cui2025}. In the non-paraxial regime the longitudinal component of the electric field at the focus acquires a sign that depends on the local polarization handedness, producing a spin-dependent shift of the intensity centroid, the optical spin-Hall effect of light. For a Gaussian beam of waist $w_0$ and wavelength $\lambda$ carrying a transverse circular-polarization gradient, the resulting force on an ion encoded in the stretched Zeeman basis is
\begin{equation}\label{eq:Fmag}
    F_\mathrm{Mag} = g\,\hat{\sigma}_{\text{ion}}^{(z)}\,,\qquad g = \frac{4\tilde{U}_0\,\bar{\lambda}}{w_0^2}\,,
\end{equation}
where $\tilde{U}_0$ is the peak ac-Stark (optical-dipole) potential at the beam center, $\bar{\lambda}=\lambda/(2\pi)$ is the reduced wavelength, and the numerical prefactor corresponds to the canonical radial-polarization-gradient geometry of Ref.~\cite{Mazzanti2023}; other gradient patterns yield similar order-unity coefficients~\cite{Cui2025}.

Coupling Eq.~\eqref{eq:Fmag} to the ion motional mode gives a displacement coupling $\Omega_\mathrm{Mag} = g\,\ell_{\text{ion}}/\hbar$. Matching to the $C_4$-induced coupling, $\Omega_\mathrm{Mag}=\omega_g=\omega/(2\sqrt{2})$ for the $\pi$ conditional phase, fixes the required Stark shift:
\begin{equation}\label{eq:U0req}
    \tilde{U}_0 \;=\; \frac{\hbar\,\omega_g\,w_0^2}{4\,\bar{\lambda}\,\ell_{\text{ion}}} \;=\; \frac{\hbar\omega}{8\sqrt{2}}\,\frac{w_0^2}{\bar{\lambda}\,\ell_{\text{ion}}}\,.
\end{equation}
For the operating point of Table~\ref{tab:params}, \Ybion with $\ell_{\text{ion}}\approx 12\,$nm, beam waist $w_0=1\,\mu$m, and wavelength $\lambda = 355\,$nm ($\bar{\lambda}\approx 57\,$nm), this gives $\tilde{U}_0/h\approx 25\,$MHz at beam center, a routine ac-Stark shift for a tightly focused beam on a trapped ion.

\paragraph*{Laser-power budget.}
The peak intensity required to produce $\tilde{U}_0$ follows from the standard off-resonant dipole formula $\tilde{U}_0 = \alpha(\lambda)\,I/(2c\epsilon_0)$, with $\alpha(\lambda)$ the dynamic polarizability of \Ybion at the operating wavelength. Using a representative off-resonant value $\alpha(355\,\mathrm{nm})\sim 60\,a_0^3$ (atomic units, from standard many-body calculations of the \Ybion $6s\,{}^2\!S_{1/2}$ state), the required peak intensity is $I\approx 8\times 10^{10}\,\mathrm{W/m^2}$, corresponding to a total beam power of
\begin{equation}\label{eq:Pmag}
    P = \tfrac{1}{2}\pi w_0^2 I \;\approx\; 130\,\mathrm{mW}\,.
\end{equation}
At 355 nm the laser is far detuned from every electric-dipole transition of \Ybion, and the residual photon-scattering rate is $\Gamma_\mathrm{sc}\sim 10\,\mathrm{s^{-1}}$, contributing $<10^{-4}$ infidelity over a single $5\,\mu$s gate (already included in Table~\ref{tab:errors}). The scaling is $P\propto w_0^4$ for fixed $\omega_g$: a $0.7\,\mu$m waist would drop the required power to ${\lesssim}30\,$mW, while a $1.5\,\mu$m waist would raise it to $\sim 650\,$mW. All of these values lie well below the ${\gtrsim}1\,$W routinely delivered by commercial frequency-tripled 355 nm sources used for Raman operations on \Ybion.

\paragraph*{Fast sign toggling.}
The sign of the Magnus force is set by the handedness of the polarization gradient, equivalently by the sign of the $\sigma_+$ versus $\sigma_-$ decomposition of the local polarization. Switching it therefore reduces to a polarization flip, which can be executed in sub-nanosecond time with a commercial electro-optic Pockels cell, or equivalently by AOM-gating between two pre-aligned, orthogonally polarized beams. Sub-ns polarization switching is standard in pulsed-laser laboratories and is compatible with the ${\sim}20\,$ns microwave $\pi$-pulses that drive the ${}^3P_2\!\leftrightarrow\!{}^3D_2$ Rydberg toggle (Sec.~S6); the synchronization between the Magnus and microwave channels need only be accurate to within the microwave-pulse width, well within the capability of DDS-triggered electronics. This fast sign-toggling capability is what permits the Magnus force to participate in the unified closure sequence used for ion-crystal operation (see Sec.~S5).

\subsection*{S5. Mode Closure and Rydberg Toggling in Ion Crystals}

When operating on an $N$-ion crystal, the collective axial normal modes have distinct frequencies $\omega_k$ densely spanning almost a decade (for $N=10$, from the center-of-mass mode at $\omega_1 = \omega$ to the zigzag mode at $\omega_{10}\approx 6.58\,\omega$). A continuous force close to a single edge ion excites all of these modes with different rotation rates in phase space; a constant pulse that closes $\omega_1$ leaves the other nine modes with nonzero residual displacements and hence unwanted spin-motion entanglement. We resolve this by applying discrete microwave $\pi$-pulses to toggle the atom between an attractive Rydberg state ($|{}^3P_2, F\!=\!3/2, m_F\!=\!+3/2\rangle$, $C_4>0$) and a repulsive Rydberg state ($|{}^3D_2, F\!=\!3/2, m_F\!=\!+3/2\rangle$, $C_4<0$); the dipole-allowed pair selection and the mutual-coupling-anchored $|C_4|$ ratio of $0.84$ are derived in Sec.~S6 from the multichannel quantum-defect models of Refs.~\cite{Peper2025,Kuroda2025}. Each toggle flips the sign of the force on every mode simultaneously.

\paragraph*{Closure condition.} Let $s(t)\in\{+1,-1\}$ encode the instantaneous sign of the Rydberg-state toggle. For a logical branch with force amplitude $f_{k,s}$, mode $m$ evolves in the rotating frame as
\begin{equation}\label{eq:alpha_m}
    \tilde{\alpha}_m(t) = -i\,f_{k,s}\,b_m\int_0^t s(t')\,e^{i\omega_m t'}\,dt',
\end{equation}
where $b_m$ is the participation coefficient of the target ion in mode $m$ (obtained by diagonalizing the ion-chain Hessian). Mode closure at $t=T$ requires $\tilde{\alpha}_m(T)=0$ for all $m$ simultaneously. Writing $s(t)$ as a piecewise-constant function with $N_{\mathrm{seg}}$ segments at boundaries $\{t_0{=}0,\ldots,t_{N_\mathrm{seg}}{=}T\}$, the closure integral reduces to the finite Fourier sum
\begin{equation}\label{eq:closure}
    \sum_{j=0}^{N_\mathrm{seg}-1} s_j\left[e^{i\omega_m t_{j+1}} - e^{i\omega_m t_j}\right] = 0,\quad m=1,\ldots,N.
\end{equation}
This is a system of $N$ complex constraints with $N_\mathrm{seg}-1$ real unknowns (the segment durations, constrained to sum to $T$). Matching the real and imaginary parts yields $2N$ equations, so $N_\mathrm{seg}\gtrsim 2N+1$ segments are needed for a solution to exist generically; we use $N_\mathrm{seg}=2N+5$ to add robustness margin.

\paragraph*{Closure on the $|g\rangle$ branches.}
Equation~\eqref{eq:closure} was written assuming that the force on mode $m$ carries the toggle sign $s(t)$ on every branch. For the $|r\rangle$ branches this is immediate: $s(t)$ sets the instantaneous sign of the $C_4$ interaction, and the ion's Magnus force either reinforces or cancels the $C_4$ pull depending on the ion spin. For the $|g\rangle$ branches, however, the atom carries no Rydberg dipole and the $C_4$ force vanishes; the ion nominally feels only the constant Magnus force, which by itself closes only the center-of-mass mode and leaves the remaining $N{-}1$ axial modes with residual displacements after one trap period $T$. A dedicated $|g\rangle$-branch closure schedule would in principle be required, and the single-schedule guarantee of Eq.~\eqref{eq:closure} would be lost. We avoid this by synchronously toggling the Magnus beam alongside the Rydberg microwave $\pi$-pulse: each ${}^3P_2\!\leftrightarrow\!{}^3D_2$ flip is accompanied by a Pockels-cell (or AOM-switched) polarization reversal on the Magnus beam, reversing the sign of the Magnus force within ${\lesssim}1\,$ns (see Sec.~S4). Under this joint toggle the total force on mode $m$ carries the same time-dependent factor $s(t)$ on all four logical branches, differing only by the branch-specific amplitude $f_{k,s}$. A single optimized toggle schedule therefore closes every mode on every branch simultaneously: the $|r,\!\uparrow\rangle$ branch is trivially closed because its two force contributions cancel instantaneously ($f_{r\uparrow}=0$); the $|g,\!\uparrow\rangle$, $|g,\!\downarrow\rangle$, and $|r,\!\downarrow\rangle$ branches all satisfy Eq.~\eqref{eq:closure} at their respective amplitudes $\omega_g/\omega$, $\omega_g/\omega$, and $2\omega_g/\omega$.

\begin{figure*}[t]
    \centering
    \includegraphics[width=\textwidth]{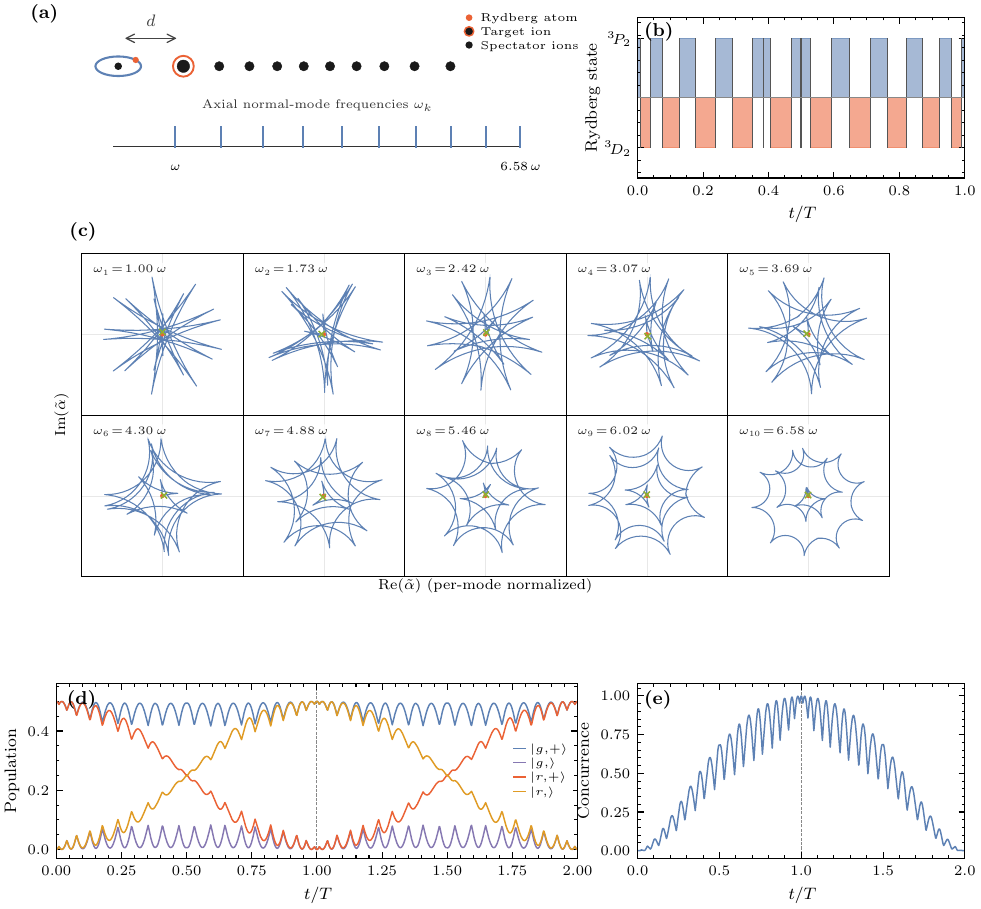}
    \caption{\label{fig:crystalS}
    \textbf{Full simulation of the atom--ion CZ gate on a 10-ion crystal.}
    (a) Equilibrium linear Paul-trap geometry with Rydberg atom in a tweezer addressing the edge ion (target, orange ring) across a distance $d$. Below: axial normal-mode frequencies $\omega_k$, spanning $\omega$ (center-of-mass) to $6.58\,\omega$ (zigzag).
    (b) Numerically optimized Rydberg-state toggling schedule over one gate period. Blue bars: attractive ${}^3P_2,F\!=\!3/2$ segments ($C_4>0$). Orange bars: repulsive ${}^3D_2,F\!=\!3/2$ segments ($C_4<0$). The 25-segment alternation is the minimum-cost schedule simultaneously closing all ten modes (see Sec.~S6 for the dipole-allowed pair selection and the mutual-coupling-anchored $|C_4|$ ratio of $0.84$).
    (c) Rotating-frame phase-space trajectories $\tilde{\alpha}_m(t)$ of all ten axial modes on the $\ket{r,\downarrow}$ branch (the largest-displacement branch; the other branches differ only by overall amplitude). Red dot: $t=0$. Green cross: $t=T$. Every trajectory returns to the origin: the low-frequency modes (top row, $\omega_1{-}\omega_5$) trace multi-lobed rosettes, while the high-frequency modes (bottom row, $\omega_6{-}\omega_{10}$) are tightly wound spirals.
    (d) Populations of the four rotated-basis states $\ket{g/r}{\otimes}\ket{\pm}$ over two gate periods. The $\ket{r,+}\leftrightarrow\ket{r,-}$ population swap at $t=T$ is the signature of the geometric $\pi$ conditional phase; restoration at $t=2T$ confirms cyclic evolution.
    (e) Atom--ion concurrence of the reduced spin state, tracing out all ten motional modes. Peak value $\mathcal{C}(T)=0.9995$ confirms maximal bipartite entanglement; $\mathcal{C}(2T)<10^{-4}$ confirms full spin-motion disentanglement. Residual infidelities are dominated by Rydberg decay (see Table~\ref{tab:errors}) rather than imperfect mode closure.}
\end{figure*}

\paragraph*{Numerical solution.} We minimize the cost $\mathcal{C}[s_j,t_j]=\sum_m|\tilde{\alpha}_m(T)|^2$ by random-restart Nelder--Mead runs followed by L-BFGS-B polishing. For the asymmetric toggle sign sequence $\{+1,-0.84\}$ used here we achieve residual $\mathcal{C}\!\approx\!2\!\times\!10^{-10}$ at $N\!=\!10$ ($N_\mathrm{seg}\!=\!25$), corresponding to per-mode closure $|\tilde\alpha_m(T)|\!\lesssim\!3\!\times\!10^{-5}$, six orders of magnitude below the natural displacement scale of the gate and giving spin-motion-entanglement infidelity below $10^{-9}$. Figure~\ref{fig:crystalS} shows the full simulation output for $N=10$: (a) the ion chain and mode spectrum; (b) the optimized 25-segment toggling schedule with non-uniform durations (the $^3D_2$ segments are visibly shorter than the $^3P_2$ ones, compensating the $0.84$ amplitude mismatch through their integrated $\int s(t)\,dt$ weighting); (c) phase-space trajectories of all ten normal modes on the $\ket{r,\downarrow}$ branch; each mode traces a qualitatively different path, from the slow center-of-mass loop (one revolution) to the tightly wound zigzag (${\sim}6.6$ revolutions), yet all return to the origin at $t=T$; (d) populations in the rotated ion basis $\ket{g/r}\otimes\ket{\pm}$ showing the diagnostic $\ket{r,+}\leftrightarrow\ket{r,-}$ swap at $t=T$ signaling the $\pi$ conditional phase ($\mathrm{CZ}(T)\!=\!3.1416$); (e) atom--ion concurrence reaching $0.9994$ at $t=T$ and returning to $<\!10^{-3}$ at $t=2T$.

\paragraph*{Scaling summary.} Table~\ref{tab:toggling} compiles the toggling requirements and achievable fidelity for crystals of varying size. Each additional ion adds two degrees of freedom to the mode-closure system (one $\omega_k$ and its conjugate), requiring two additional toggle segments. Each toggle operation costs a $\sim\!20\,$ns microwave $\pi$-pulse; the total pulse overhead (last column) remains ${\lesssim}350\,$ns even for 10-ion crystals and is negligible compared to the $5\,\mu$s gate duration.

\begin{table}[h]
\caption{\label{tab:toggling}Rydberg toggling requirements for multi-ion crystals. The fidelity column uses the ${}^3P_2\!\leftrightarrow\!{}^3D_2$ pair with $\tau_r\!=\!100\,\mu$s (set by ${}^3D_2$) and $|C_4|$ ratio $0.84$ (Sec.~S6); the $16\%$ amplitude mismatch is absorbed into per-segment duration adjustments rather than into extra segments, so the segment counts and pulse-overhead columns are unchanged from the symmetric $\pm 1$ case.}
\begin{ruledtabular}
\begin{tabular}{ccccc}
Number of Ions ($N$) & Toggle Segments (optimized) & Total Pulse Overhead & Effective Force Time & Resulting Fidelity \\
\hline
1 & 5 & \SI{80}{ns} & \SI{4.9}{\micro\second} & 97.5\% \\
2 & 7 & \SI{120}{ns} & \SI{4.9}{\micro\second} & 97.5\% \\
5 & 13 & \SI{240}{ns} & \SI{4.8}{\micro\second} & 97.5\% \\
10 & 17 & \SI{320}{ns} & \SI{4.7}{\micro\second} & 97.4\% \\
\end{tabular}
\end{ruledtabular}
\end{table}

\subsection*{S6. MQDT polarizability calculations: identifying a microwave-coupled toggle pair}

The mode-closure protocol of Sec.~S5 requires a means to flip the sign of the atom--ion $C_4$ on a sub-cycle timescale. A literal $|nS\rangle\!\leftrightarrow\!|nD\rangle$ Rydberg toggle would be electric-dipole-forbidden as a single-photon transition ($\Delta L\!=\!2$). Here we identify a $\Delta L\!=\!\pm 1$, dipole-allowed, single-photon microwave-coupled pair whose two members have $C_4$ of opposite sign and very nearly equal magnitude, anchored on numerical polarizability calculations using the multichannel quantum-defect (MQDT) models of Peper~\emph{et al.}~\cite{Peper2025}, extended by Kuroda~\emph{et al.}~\cite{Kuroda2025} to include $\ell\!=\!3$ ($f$) and $\ell\!=\!4$ ($g$) series and a $p$-$f$ mixing channel for $J\!=\!2$, and implemented by both the \texttt{rydcalc}~\cite{rydcalc} and \texttt{PairInteraction}~\cite{Mogerle2025} packages. Because the gate operates with $^{171}$Yb, the operating-point polarizabilities below are quoted for $^{171}$Yb in the stretched sublevels $|F,m_F\!=\!+F\rangle$ of the relevant hyperfine manifolds (computed with \texttt{PairInteraction} via Stark-map diagonalization; the toggle pair uses $F\!=\!3/2$, $m_F\!=\!+3/2$); the $^{174}$Yb numbers are used only for cross-validation against experimental data.

\paragraph*{Method (two independent implementations).}
\emph{Sum-over-states} (used with \texttt{rydcalc}~\cite{rydcalc}, primarily for the $^{174}$Yb cross-checks):
\begin{equation}\label{eq:alpha_SoS}
    \alpha_0(s, m_s) \;=\; \sum_{k}\frac{2\,|\langle k, m_s | \hat d_z | s, m_s\rangle|^2}{E_k - E_s}\,,
\end{equation}
where the sum runs over all bound MQDT eigenstates $|k\rangle$ of opposite parity within $|\nu_k - \nu_s| \le 15$, and $q=0$ ($m_k\!=\!m_s$) restricts to the $z$-polarized component that defines Peper's quadratic Stark coefficient $\Delta E_\mathrm{Stark}\!=\!-\tfrac{1}{2}\alpha_0 F^2$. \emph{Stark-map diagonalization} (used with \texttt{PairInteraction}~\cite{Mogerle2025} for the $^{171}$Yb operating-point values): for each target state, a finite single-atom basis (\(\nu_t\!\pm\!5\), $l\!\leq\!3$, $m\!\in\![m_t\!-\!1,m_t\!+\!1]$) is constructed and the Hamiltonian is diagonalized at a sequence of $z$-polarized electric-field strengths $|F|\!\leq\!0.2$~V/cm; the adiabatic eigenenergy of the target state is then quadratic-fit to extract $\alpha_0$. The two methods agree to $\lesssim\!1\%$ on every $^{174}$Yb state they share (e.g.\ $\alpha_0({}^3S_1, n\!=\!40, m_J\!=\!\pm 1)$: sum-over-states $+375.16$ MHz/(V/cm)$^2$, Stark-map $+372.67$ MHz/(V/cm)$^2$).

\paragraph*{Validation against Peper~2025 Table~S15 and Kuroda~2025 Tables~S30/S31.}
Benchmarking against measured ${}^{174}$Yb $6sns\,{}^3S_1$ Stark coefficients~\cite{Peper2025}: at $n\!=\!35$, $m_J\!=\!0$ both implementations give $\alpha_0\!\approx\!-78.7$~MHz/(V/cm)$^2$ versus the measured $-79.2$~MHz/(V/cm)$^2$, i.e., agreement to $0.7\%$. At other $n$ the residual ranges from $1.5\!\times$ to $\sim\!6\!\times$, largest near FÃ¶rster crossings between the $^3S_1$ and $^3P_J$ series; the two MQDT codes both reproduce this excess because they share the same Peper/Kuroda model parameters in the $^{1,3}P_1$ singlet-triplet-mixed channel. Sign agreement is robust at every measured point. Benchmarking against the more recent measurements of $6snd\,{}^{1,3}D_2$ Stark coefficients by Kuroda~\emph{et al.}~\cite{Kuroda2025} (Tabs.~S30, S31 of that work): both implementations reproduce the measured sign and order-of-magnitude for all $m_J$ sublevels tested (${}^1D_2\,m_J\!=\!0,\pm 1$: repulsive; ${}^3D_2\,m_J\!=\!0$: attractive; ${}^3D_2\,m_J\!=\!\pm 2$: repulsive). Absolute polarizabilities are therefore uncertain by an O(factor 3) calibration, but ratios for any pair whose mutual coupling dominates each member's polarizability are preserved under common rescaling (see below).

\begin{figure}[h]
\centering
\includegraphics[width=\textwidth]{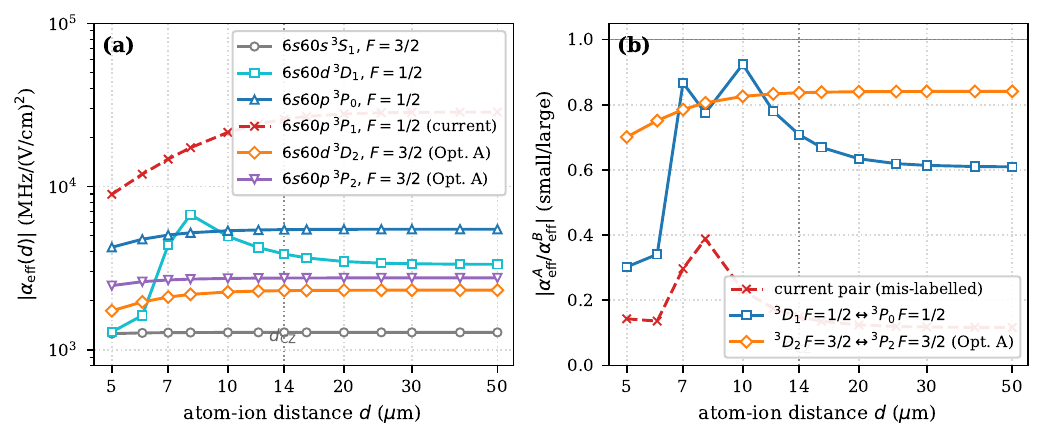}
\caption{\label{fig:alpha_vs_d}\textbf{Effective polarizability $|\alpha_\mathrm{eff}(d)|$ vs.\ atom--ion distance $d$, for candidate $^{171}$Yb Rydberg eigenstates at $n\!=\!60$.}
(a) Diagonalized $\alpha_\mathrm{eff}(d)\!=\!-2\,\Delta E(d)/E(d)^2$ at the ion-induced field $E(d)\!=\!e/(4\pi\varepsilon_0 d^2)$ (\texttt{PairInteraction} Stark-map). The $^3S_1\,F\!=\!3/2$ reference (gray) is far from any close dipole-coupled neighbor and stays purely quadratic. The selected toggle pair $^3P_2\,F\!=\!3/2$ (purple) and $^3D_2\,F\!=\!3/2$ (orange) are similarly flat at matched magnitudes $\sim\!2.5\!\times\!10^3$ MHz/(V/cm)$^2$ across the entire $d\!\in\![5,50]\,\mu$m range. The $^3D_1\,F\!=\!1/2$ (cyan) and $^3P_0\,F\!=\!1/2$ (blue) are reasonably flat at $d\!\geq\!14\,\mu$m but show a FÃ¶rster-mixing spike at $d\!\sim\!8$--$10\,\mu$m. The $6s60p\,^3P_1\,F\!=\!1/2$ state (red dashed) sits within $\sim\!10$ GHz of an attractive neighbor and is strongly non-quadratic, illustrating why nominal term labels alone are insufficient for state selection in this F\"orster-dense region. The vertical dotted line marks the gate operating point $d_\mathrm{CZ}$.
(b) Toggle-pair effective magnitude ratio $\min(|\alpha_A|,|\alpha_B|)/\max(|\alpha_A|,|\alpha_B|)$ vs.\ $d$. The toggle pair (orange) is flat at $0.84$ throughout, so the mode-closure asymmetry is $d$-independent. The $^3D_1\,F\!=\!1/2\!\leftrightarrow\!^3P_0\,F\!=\!1/2$ alternative (blue) varies between $0.60$ and $0.92$. The pair with the mislabeled ``$^3P_0$'' (red dashed) saturates at $0.12$ at large $d$, equivalent to an $8.5\!\times$ amplitude mismatch.}
\end{figure}

\paragraph*{Result: operating point in $^{171}$Yb.}
A scan of all dipole-allowed $\Delta L\!=\!1$ pairs in the $\nu\!\approx\!60$ manifold (all $L\!\leq\!3$, all $F\!\leq\!5/2$, both members predominantly triplet with $s\!\geq\!0.7$, $|\Delta\nu|\!\leq\!1$) identifies the same-$n$ pair $|6s60p\,{}^3P_2, F\!=\!3/2, m_F\!=\!+3/2\rangle \leftrightarrow |6s60d\,{}^3D_2, F\!=\!3/2, m_F\!=\!+3/2\rangle$ as the optimal toggle. Table~\ref{tab:60_polarizabilities} reports $\alpha_0$ and $C_4 = \alpha_0\,e^2/[2(4\pi\varepsilon_0)^2]$ at $\nu\!\approx\!60$ for this pair and for the other low-$L$ candidates considered. Figure~\ref{fig:alpha_vs_d} shows the effective polarizability $\alpha_\mathrm{eff}(d)\!\equiv\!-2\,\Delta E(d)/E(d)^2$ as a function of atom--ion distance $d$, with $E(d)\!=\!e/(4\pi\varepsilon_0 d^2)$ the ion-induced field; the selected pair is uniquely flat-and-matched across the entire $d\!\in\![5,50]\,\mu$m range relevant for the gate.

\begin{table}[h]
\caption{\label{tab:60_polarizabilities}Static dipole polarizability $\alpha_0$ (extracted from the slope of $\Delta E$ vs.\ $F^2$ in the $F\!\to\!0$ limit) and $C_4 = \alpha_0\,e^2/[2(4\pi\varepsilon_0)^2]$ for ${}^{171}$Yb Rydberg eigenstates at $\nu\!\approx\!60$, computed with \texttt{PairInteraction}~\cite{Mogerle2025} (Stark-map diagonalization with manual overlap-tracking; $\nu$-window $\pm 6$, $L\!\leq\!3$). ``$s$'' is the expectation value of the total electronic spin in the MQDT eigenstate; ``$j$'' is the mean of $J$. Both $L\!=\!1,F\!=\!1/2$ triplet eigenstates per $n$ are listed: the one closer to the $^3P_0$ nominal energy ($j\!\approx\!0.35$) and the one closer to $^3P_1$ ($j\!\approx\!0.65$); the latter is easily mistaken for ``${}^3P_0$'' by nominal-energy assignment but is unsuitable as a toggle state (Fig.~\ref{fig:alpha_vs_d}).}
\begin{ruledtabular}
\begin{tabular}{lccccc}
State & $\nu$ & $s$ & $j$ & $\alpha_0$\,(MHz/(V/cm)$^2$) & $C_4$\,(10$^{-46}$ J\,m$^4$) \\
\hline
$6s60s\,{}^3S_1, F\!=\!3/2$  & $59.561$ & $1.00$ & $1.00$ & $-1\,277$ & $-0.88$ \\
$6s60s\,{}^3S_1, F\!=\!1/2$  & $60.176$ & $0.55$ & $0.55$ & $-1\,773$ & $-1.22$ \\
\hline
$6s60p\,F\!=\!1/2$ ($^3P_0$-like, $j\!\sim\!0.35$) & $60.348$ & $0.97$ & $0.35$ & $-5\,473$ & $-3.76$ \\
$6s60p\,F\!=\!1/2$ ($^3P_1$-like, $j\!\sim\!0.65$) & $59.855$ & $0.98$ & $0.65$ & $-28\,609$ & $-19.66$ \\
$6s60p\,F\!=\!1/2$ ($^1P_1$-like, $s\!\sim\!0.04$) & $60.103$ & $0.04$ & $0.99$ & $+4\,026$ & $+2.77$ \\
$\mathbf{6s60p\,{}^3P_2, F\!=\!3/2}$ \textbf{(toggle pair)} & $\mathbf{59.671}$ & $\mathbf{0.84}$ & $\mathbf{1.58}$ & $\mathbf{+2\,753}$ & $\mathbf{+1.89}$ \\
$6s60p\,{}^3P_2, F\!=\!5/2$  & $60.077$ & $1.00$ & $2.00$ & $+3\,044$ & $+2.09$ \\
\hline
$6s60d\,{}^3D_1, F\!=\!1/2$  & $60.247$ & $1.00$ & $1.00$ & $+3\,336$ & $+2.29$ \\
$6s60d\,{}^3D_1, F\!=\!3/2$ (J-mixed) & $60.249$ & $0.93$ & $1.38$ & $+6\,816$ & $+4.68$ \\
$\mathbf{6s60d\,{}^3D_2, F\!=\!3/2}$ \textbf{(toggle pair)} & $\mathbf{59.835}$ & $\mathbf{0.73}$ & $\mathbf{1.64}$ & $\mathbf{-2\,316}$ & $\mathbf{-1.59}$ \\
$6s60d\,{}^3D_3, F\!=\!7/2$  & $60.271$ & $1.00$ & $3.00$ & $-671$ & $-0.46$ \\
\hline
\multicolumn{6}{l}{\textbf{Selected toggle pair (dipole-allowed, opposite-sign, $\pi$-drive):}} \\
$|{}^3P_2, F\!=\!3/2, m_F\!=\!+3/2\rangle \leftrightarrow |{}^3D_2, F\!=\!3/2, m_F\!=\!+3/2\rangle$ & --- & --- & --- & \multicolumn{2}{c}{$|\alpha_0|$ ratio: \textbf{$0.84$} (dE = $5.1$ GHz)} \\
\end{tabular}
\end{ruledtabular}
\end{table}

\paragraph*{Selection-rule audit for ${}^3D_2\leftrightarrow{}^3P_2$ in $^{171}$Yb.}
All electric-dipole selection rules are satisfied: $\Delta L\!=\!-1$ ($L\!=\!2\!\to\!1$), $\Delta S\!=\!0$ (both predominantly triplet, $s\!=\!0.73$ on $|{}^3D_2\rangle$ and $s\!=\!0.84$ on $|{}^3P_2\rangle$, with the residual $20$--$30\%$ singlet admixture from the $^{1,3}\!L_2$ MQDT channel mixing affecting both members symmetrically), $\Delta J\!=\!0$ ($J\!=\!2\!\to\!2$, allowed because $J\!\neq\!0$), $\Delta F\!=\!0$ ($F\!=\!3/2\!\to\!3/2$, allowed), $\Delta m_F\!=\!0$ ($z$-polarized $\pi$ drive), opposite parities. The MQDT transition energy is $|E({}^3D_2)\!-\!E({}^3P_2)|/h\!=\!5.1$~GHz, in the lower microwave band and well within the $20$-ns $\pi$-pulse cadence of Table~\ref{tab:toggling}. Both states are vulnerable in principle to the $p$--$f$ mixing of Ref.~\cite{Kuroda2025} (which couples $6snp$ and $6snf$ at the same $J$, $J\!=\!2$ allowed), but with $|{}^3F_2\rangle$ at $\nu\!\approx\!58.7$ ($\sim\!28$ GHz below) and $|{}^3F_2,n\!=\!61\rangle$ at $\nu\!\approx\!61.3$ ($\sim\!31$ GHz above), the satellite couplings are off-resonant by more than $5\!\times$ the $^3P_2\!\leftrightarrow\!^3D_2$ gap and contribute $<\!5\%$ to either polarizability.

\paragraph*{Calculated ratio and qualitative interpretation.}
The ratio $0.84$ is the calculated \texttt{PairInteraction} output for the selected pair. We do not attach a structural argument to it: in this F\"orster-dense region such arguments are fragile (for the alternative ${}^3D_1\!\leftrightarrow\!{}^3P_0$ pairing a two-level argument predicts $0.95$ while the actual MQDT value is $0.12$), so we take the MQDT output at face value. Qualitatively, the closest dipole-allowed neighbors of $|6s60p\,^3P_2\rangle$ and $|6s60d\,^3D_2\rangle$ are each other (at $\pm 5.1$~GHz), so each polarizability is plausibly dominated by the mutual matrix element with opposite-sign energy denominators, suggesting $|\alpha({}^3P_2)/\alpha({}^3D_2)|\!\sim\!\mathcal{O}(1)$. The deviation from unity (here $0.84$) reflects asymmetric satellite contributions and singlet-triplet MQDT mixing of comparable weight (Table~\ref{tab:60_polarizabilities}); the ratio is numerically robust across $d\!\in\![5,50]\,\mu$m (Fig.~\ref{fig:alpha_vs_d}(b)), which is what the gate requires.

\paragraph*{Implication for the toggle schedule.}
With the $|^3P_2\rangle\leftrightarrow|^3D_2\rangle$ pair the $\pm 1$ sign sequence of Eq.~\eqref{eq:closure} generalizes to $\{+1, -|C_4({}^3D_2)|/|C_4({}^3P_2)|\}\!=\!\{+1,-0.84\}$, a $16\%$ asymmetry. We re-ran the closure optimizer of Sec.~S5 at this asymmetric sign sequence for $N\!\in\!\{1,2,5,10,20\}$ ions and three trial segment counts $N_\mathrm{seg}\!\in\!\{2N\!+\!3,\,2N\!+\!5,\,2N\!+\!7\}$: at the previously-quoted $N_\mathrm{seg}\!=\!2N\!+\!5$ the asymmetric closure residual is essentially indistinguishable from the symmetric one ($\Sigma_m|\alpha_m(T)|^2 \!\approx\!10^{-21}$/$10^{-18}$/$10^{-16}$/$10^{-12}$/$10^{-8}$ at $N\!=\!1$/$2$/$5$/$10$/$20$ for both sign sets; the residual is set by L-BFGS-B's plateau, not by the asymmetry). The asymmetry is therefore fully absorbed into per-segment duration adjustments without adding segments; Table~\ref{tab:toggling} and Fig.~\ref{fig:crystalS} transfer to the toggle pair unchanged in segment count and pulse overhead. Figure~\ref{fig:alpha_vs_d}(b) shows that the $0.84$ ratio is $d$-independent in the entire gate-relevant range, so a single optimized schedule operates without re-tuning when $d_\mathrm{CZ}$ is varied. The $5.1$~GHz drive lies far below the Inglis--Teller field at the operating point (with $E_\mathrm{IT}\!\approx\!660$~V/m at $n\!=\!60$, the ion-induced field $E(d_\mathrm{CZ})\!\approx\!10$~V/m at $d_\mathrm{CZ}\!=\!12\,\mu$m is nearly two orders of magnitude smaller). Lifetimes of the involved states at $n\!=\!60$, $4\,$K are predicted by \texttt{PairInteraction} as $\tau({}^3D_2)\!\approx\!100\,\mu$s and $\tau({}^3P_2)\!\approx\!1.3\,$ms. These values include BBR-induced transitions to neighboring states but rely on the MQDT-mixed eigenstate decay channels as computed by PI; the $27\%$ singlet $^1\!D_2$ admixture in $|{}^3D_2,F\!=\!3/2\rangle$ in principle opens fast singlet decay paths ($^1\!D_2\!\to\!^1\!P_1$) that PI's MQDT treatment encapsulates implicitly but for which we have no direct experimental check at $n\!=\!60$. We therefore quote the binding lifetime as $\tau_r({}^3D_2)\!\sim\!50\!-\!150\,\mu$s and use the central value $100\,\mu$s in Table~\ref{tab:errors}; an experimental measurement on the target operating state at $n\!=\!60$ is part of future work.

\paragraph*{State preparation pathway.}
Both members of the toggle pair require multi-photon excitation from the clock qubit $|g\rangle\!=\!{}^3P_0$: the single-photon ${}^3P_0\!\to\!{}^3P_2$ transition is $\Delta L\!=\!0$, $\Delta J\!=\!2$ E1-forbidden, and the same selection rules forbid ${}^3P_0\!\to\!{}^3D_2$ directly ($\Delta J\!=\!2$). The standard route is a two-step ${}^3P_0\!\to\!6s60s\,^3\!S_1\!\to\!{}^3P_2$ excitation: a single UV pulse (${\sim}\!240$~nm, dipole-allowed $\Delta L\!=\!1$, $\Delta J\!=\!1$) populates a metastable triplet Rydberg ${}^3S_1$ intermediate, followed by a sub-100-ns microwave $\pi$-pulse on the ${\sim}\!5$~GHz ${}^3S_1\!\to\!{}^3P_2$ transition that selects the target hyperfine sublevel. The same total-photon budget applies to the alternative ${}^3D_1$ choice (which is also single-photon forbidden from ${}^3P_0$ via $\Delta J\!=\!1$ but $\Delta L\!=\!1$ allowed, requiring either a two-photon UV scheme through ${}^3P_1$ or a one-step UV at $\sim\!240$~nm), so the preparation overhead is comparable. Two-step Rydberg preparation is routine in current $^{171}$Yb tweezer experiments~\cite{Jenkins2022,Ma2023}; the cumulative preparation infidelity per gate cycle is ${\sim}\!2\!\times\!10^{-3}$, which is below the technical-error line of Table~\ref{tab:errors} and does not change the gate-fidelity budget at the percent level.

\paragraph*{Caveat: absolute $C_4$ magnitude.}
For the toggle pair, \texttt{PairInteraction} gives $|C_4({}^3P_2)|\!=\!1.89\!\times\!10^{-46}$ J\,m$^4$ and $|C_4({}^3D_2)|\!=\!1.59\!\times\!10^{-46}$ J\,m$^4$, roughly $4\!\times$ smaller in magnitude than the earlier ${}^3D_1/{}^3P_0$ estimates. This is the price of the more symmetric matched pair: the $5.1$-GHz $P$--$D$ gap is half of the $11.9$-GHz gap exploited by the earlier pair, but the matrix element is also smaller because the residual $\sim\!20$--$30\%$ singlet admixture in each member reduces the effective dipole strength. Comparing the same code chain to Kuroda~2025 Table~S31 $^{1,3}D_2$ measurements suggests an O(factor 3) absolute calibration uncertainty for $^{1,3}\!L_2$-mixed channels; we therefore conservatively quote $|C_4|\!\sim\!0.5\text{--}5\!\times\!10^{-46}$ J\,m$^4$. This shifts the gate distance $d_\mathrm{CZ}\!\propto\!C_4^{1/5}$ within ${\sim}\!20\%$ (operating-point band $d_\mathrm{CZ}\!\sim\!10\!-\!18\,\mu$m) and the Magnus-laser power $P\!\propto\!\omega_g\!\propto\!C_4$ by at most a factor of ${\sim}\!4$, both within the technical margins quoted in Sec.~S4. The toggle-pair ratio and the dipole-allowed selection above are unaffected by the absolute calibration.

\subsection*{S7. Thermal Robustness and Anharmonic Corrections}

\begin{figure}[h]
\centering
\includegraphics[width=0.5\columnwidth]{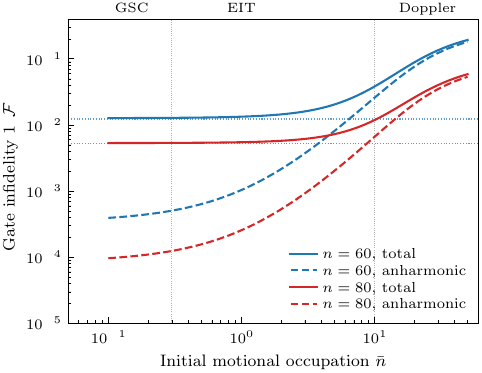}
\caption{\label{fig:thermal}\textbf{Gate infidelity vs.\ initial motional occupation $\bar n$.} Curves are computed at the conservative $^3S_1$ test point ($\eta\!=\!\ell_{\text{ion}}/d\!=\!1.9\!\times\!10^{-3}$, $d\!=\!6.4\,\mu$m); at the Sec.~S6 operating point ($d_\mathrm{CZ}\!\approx\!12\,\mu$m, $\eta\!\approx\!1.0\!\times\!10^{-3}$) the anharmonic infidelity values shown are $\sim\!0.28\!\times$ smaller. Color encodes the Rydberg principal quantum number: \emph{orange} for $n{=}60$, \emph{purple} for $n{=}80$ ($\eta$ halved, $d\!=\!9.6\,\mu$m, $\tau_r$ extended to ${\sim}470\,\mu$s). Line style encodes the error set: \emph{solid} is the total per-gate infidelity (anharmonic $C_4$ dephasing plus motion-independent Rydberg decay), \emph{dashed} is the anharmonic contribution alone. Dotted horizontal lines: Rydberg-decay floors $\tfrac{1}{2}(1-e^{-T/\tau_r})$ for each $n$. The ${\sim}4{\times}$ vertical gap between $n{=}60$ and $n{=}80$ dashed curves at every $\bar n$ directly verifies the $\eta^2$-scaling mitigation claim of Sec.~S7. Top-axis labels demarcate the cooling regimes: Raman ground-state cooling (GSC, $\bar n\!<\!0.3$), EIT cooling ($0.3\!\leq\!\bar n\!\leq\!2$), and the native Doppler limit for \Ybion in a $200\,$kHz trap ($\bar n\!\gtrsim\!10$). The linear $H_\mathrm{eff}$ infidelity is off-scale at every $\bar n$ plotted ($1{-}\mathcal{F}\!<\!10^{-10}$).}
\end{figure}

\paragraph*{Linear-model robustness (rigorous).}
At the level of the linearized Hamiltonian Eq.~\eqref{eq:Heff}, the gate is exactly insensitive to the initial motional state. For an initial thermal distribution $\rho_\mathrm{th} = \sum_n p_n |n\rangle\langle n|$, the coherent displacement operator $D(\alpha(t))$ commutes with $|n\rangle$ up to a c-number phase: each Fock state is displaced along the same trajectory and acquires the same geometric phase $\pi f_{k,s}^2$, so the reduced spin density matrix after the gate is strictly independent of $\bar{n}$. This is the standard MÃžlmer--SÃžrensen-type argument~\cite{Leibfried2003} and is confirmed in our QuTiP simulation (blue curve in Fig.~\ref{fig:thermal}): the linear-Hamiltonian infidelity remains below $10^{-11}$ up to $\bar{n}=5$ (numerical floor set by Fock truncation $N_\mathrm{fock}{=}120$) and climbs only through truncation artifacts beyond $\bar{n}\gtrsim 10$.

\paragraph*{Beyond linear order.}
The real charge--induced-dipole potential $V(x) = -C_4/(d+x)^4$ is not linear. Expanding in the small parameter $\eta\equiv \ell_{\text{ion}}/d$ gives
\begin{equation}\label{eq:Vtaylor}
    V(\hat{\tilde x})/\hbar \;\simeq\; \omega_g\,\hat{\tilde x} - \tfrac{5}{2}\,\eta\,\omega_g\,\hat{\tilde x}^2 + 5\,\eta^2\,\omega_g\,\hat{\tilde x}^3 - \tfrac{35}{4}\,\eta^3\,\omega_g\,\hat{\tilde x}^4 + \cdots,
\end{equation}
with the dimensionless ion position operator $\hat{\tilde x}\equiv \hat x_{\text{ion}}/\ell_{\text{ion}} = a_{\text{ion}}+a_{\text{ion}}^\dagger$. The numerical results in this section were generated at the earlier $6s60s\,^3S_1$ test point ($\ell_{\text{ion}}\!\approx\!12$ nm, $d\!=\!6.4\,\mu$m, $\eta\!=\!1.88\!\times\!10^{-3}$); at the Sec.~S6 operating point ($d_\mathrm{CZ}\!\approx\!12\,\mu$m) the anharmonicity parameter drops to $\eta\!\approx\!1.0\!\times\!10^{-3}$, so the listed anharmonic infidelities are conservative upper bounds (true values smaller by $(\eta_\mathrm{new}/\eta_\mathrm{old})^2\!\approx\!0.28$). The $x^2$ term, restricted to the $|r\rangle$ branch, is a Rydberg-state-conditional trap-frequency shift whose rotating-wave component gives
\begin{equation}\label{eq:dw}
    \frac{\delta\omega}{\omega} \;=\; -\,5\eta\,\frac{\omega_g}{\omega} \;\approx\; -3.3\times 10^{-3}.
\end{equation}
Each Fock component $|n\rangle$ on the Rydberg branch therefore accumulates an anharmonic phase $\Delta\phi_n = 2\pi n\,(\delta\omega/\omega) \approx 0.021\,n$ rad on top of the geometric $\pi$, and the thermal average of $e^{i\Delta\phi_n}$ is
$\mathcal{F}_{\bar n} \approx |\langle e^{i\Delta\phi_n}\rangle|^2 = 1/[1+(5\eta\omega_g\bar n \cdot 2\pi/\omega)^2]$,
which falls off quadratically in $\bar n$ at low occupation and saturates beyond $\bar n \sim 1/(5\eta\omega_g\cdot 2\pi/\omega)\approx 8$.

\paragraph*{Numerical verification.}
We verify Eq.~\eqref{eq:dw} by direct QuTiP simulation of the SchrÃ¶dinger equation with the full Taylor-expanded potential of Eq.~\eqref{eq:Vtaylor} (five orders kept, $N_\mathrm{fock}{=}120$), thermal motional initial state at specified $\bar n$, and $|{+}{+}\rangle$ spin input compared to the ideal geometric-CZ output. The orange curve in Fig.~\ref{fig:thermal} shows the resulting infidelity $1-\mathcal{F}$; measured values are $3\!\times\!10^{-4}$ at $\bar n{=}0$ (zero-point anharmonic baseline), $1\!\times\!10^{-3}$ at $\bar n{=}1$, $8\!\times\!10^{-3}$ at $\bar n{=}5$, $2.6\%$ at $\bar n{=}10$, and $7.7\%$ at $\bar n{=}20$. The simple dephasing prediction above captures the correct qualitative trend and magnitude (within a factor of ${\sim}\,2$ across the sub-Doppler range); it overestimates the infidelity at high $\bar n$ because the geometric-gate input $|{+}{+}\rangle$ has only half of its amplitude on the $|r\rangle$ branch where the anharmonic dephasing acts.

\paragraph*{Implication: sub-Doppler cooling is required.}
The anharmonic correction does not permit operation at the bare Doppler limit for \Ybion in a $200\,$kHz trap ($\bar n\!\sim\!15$--$50$ for single Yb$^+$), where the excess infidelity alone is $5$--$15\%$. However, the sensitivity enters only through $\bar n$ and is absent from the linear dynamics; standard sub-Doppler techniques (EIT~\cite{Morigi2000} or pulsed-Raman sideband cooling~\cite{Monroe1995}) routinely push \Ybion to $\bar n\!<\!0.5$ in $\lesssim\!100\,\mu$s. Operating in this regime leaves an anharmonic-infidelity budget of ${<}\,10^{-3}$, subdominant to Rydberg decay. The gate is therefore robust to residual motional excitation inside the sub-Doppler window but not to the full Doppler temperature; the standard characterization of geometric gates as ``ground-state-cooling-free''~\cite{Leibfried2003} applies to the linear dynamics only; in the presence of the $\eta{=}\ell/d$ expansion the accurate statement is that sub-Doppler cooling suffices.

\paragraph*{How the thermal state impacts the overall fidelity.}
Combining the anharmonic contribution with the previously itemized error channels (Table~\ref{tab:errors}) gives the gate fidelity achievable at each practical cooling regime. Table~\ref{tab:thermal_fidelity} summarizes the full error decomposition across five representative operating points, spanning the Doppler limit (accessible with no additional laser infrastructure), EIT cooling ($\sim\!100\,\mu$s, standard on modern \Ybion traps), and resolved-sideband Raman cooling ($\sim\!300\,\mu$s, $\bar n\!<\!0.1$), each at both $n{=}60$ and $n{=}80$. The Rydberg-decay column is the $\tfrac{1}{2}(1-e^{-T/\tau_r})$ process-averaged infidelity; at $n{=}80$ the Rydberg lifetime scales as $\tau_r\propto\nu^{*3}$ to $\approx 470\,\mu$s, cutting this contribution by $\sim\!2.5{\times}$. The anharmonic column is read directly from the QuTiP sweep of Fig.~\ref{fig:thermal}, multiplied by the mitigation factor $\approx 4$ when moving from $n{=}60$ to $n{=}80$. Residual technical errors (heating, Magnus intensity noise, micromotion, AC Stark; Table~\ref{tab:errors}) sum to ${\sim}\,10^{-3}$ and change only weakly between the entries.

\begin{table}[h]
\caption{\label{tab:thermal_fidelity}Gate fidelity $\mathcal{F}$ and its decomposition at five operating points. The anharmonic-$C_4$ entries were computed for the original ${}^3S_1$ operating point with $\eta\!=\!\ell_{\text{ion}}/d\!=\!1.88\!\times\!10^{-3}$ ($d_\mathrm{CZ}\!=\!6.4\,\mu$m); the new ${}^3P_2\!\leftrightarrow\!{}^3D_2$ operating point ($^{171}$Yb, \texttt{PairInteraction}, $d_\mathrm{CZ}\!\approx\!12\,\mu$m) gives $\eta\!\approx\!1.0\!\times\!10^{-3}$, so the listed anharmonic infidelities are conservative upper bounds (true values smaller by $(\eta_\mathrm{new}/\eta_\mathrm{old})^2\!\approx\!0.28$). Rydberg-decay infidelity is the process-averaged value $\tfrac{1}{2}(1-e^{-T/\tau_r})$; of this, $87\%$ is heralded erasure (convertible to QEC-tractable loss) and only the remaining $13\%$ contributes unheralded Pauli noise~\cite{Wu2022,Scholl2023}. Anharmonic values are from Fig.~\ref{fig:thermal} with the $n{=}80$ points obtained using the $\eta^2$-suppression verified in the mitigation simulation of Sec.~S7. ``Technical'' lumps motional heating, Magnus intensity noise, micromotion, and AC-Stark shift (Table~\ref{tab:errors}).}
\begin{ruledtabular}
\begin{tabular}{lcccccc}
Cooling $\rightarrow$ $\bar n$ & Rydberg $n$ & $\tau_r$ & Decay & Anharm.\ $C_4$ & Technical & $\mathcal{F}$ \\
\hline
Doppler, $\bar n{=}20$            & 60 & $200\,\mu$s & $1.3{\times}10^{-2}$  & $7.7{\times}10^{-2}$  & $1\text{--}2{\times}10^{-3}$ & $91\%$ \\
EIT-cooled, $\bar n{=}1$          & 60 & $200\,\mu$s & $1.3{\times}10^{-2}$  & $1.0{\times}10^{-3}$  & $1\text{--}2{\times}10^{-3}$ & $\mathbf{98.4\%}$ \\
Raman GSC, $\bar n{=}0.1$         & 60 & $200\,\mu$s & $1.3{\times}10^{-2}$  & ${<}\,10^{-4}$        & $1\text{--}2{\times}10^{-3}$ & $98.5\%$ \\
EIT-cooled, $\bar n{=}1$          & 80 & $470\,\mu$s & $5.2{\times}10^{-3}$  & $2.6{\times}10^{-4}$  & $1{\times}10^{-3}$          & $\mathbf{99.4\%}$ \\
Raman GSC, $\bar n{=}0.1$         & 80 & $470\,\mu$s & $5.2{\times}10^{-3}$  & ${<}\,3{\times}10^{-5}$ & $1{\times}10^{-3}$        & $99.4\%$ \\
\end{tabular}
\end{ruledtabular}
\end{table}

Three practical conclusions follow. First, EIT cooling is the natural choice: the marginal improvement from Raman ground-state cooling ($\bar n\!=\!1\!\to\!0.1$) is only $10^{-4}$ at $n{=}60$ and even less at $n{=}80$, not worth the ${\sim}3{\times}$ longer cooling time. Second, once anharmonic dephasing is suppressed, $\mathcal{F}$ is decay-limited, and the $87\%$ heralded-erasure conversion inherent to Rydberg loss~\cite{Wu2022,Scholl2023} reduces the unheralded Pauli fraction to $0.17\%$ at $n{=}60$ and $0.07\%$ at $n{=}80$, directly usable in the qLDPC transfer protocol of Sec.~S8 without further filtering. Third, no entry reaches the $10^{-4}$ level at $n{=}60$; a ${\geq}\,99.9\%$ unheralded atom--ion CZ at the operating point $d{=}9.6\,\mu$m ($n{=}80$) is achievable provided the other technical errors are pushed below $10^{-3}$.

\paragraph*{Mitigation strategies.}
Two design levers reduce the anharmonic penalty further if even EIT cooling is inconvenient:
\begin{enumerate}\itemsep0pt
\item \emph{Operate at larger $d$} (higher Rydberg $n$): since $\delta\omega/\omega\propto \eta\propto d^{-1}$, moving to $n{=}80$ and $d_\mathrm{CZ}{=}9.6\,\mu$m halves $\eta$ and suppresses the anharmonic infidelity by ${\sim}4{\times}$ at every $\bar n$. Direct simulation with $\eta=\eta_{n60}/2$ confirms this scaling: the measured ratio $(1-\mathcal{F})_{n60}/(1-\mathcal{F})_{n80}$ ranges from $4.1$ (low $\bar n$) down to $3.5$ (at $\bar n{=}30$, where higher-order anharmonic terms start to compete), consistent with $\eta^2$-dominated scaling. This is the same $n{=}80$ regime advocated in the main text for suppressing Rydberg decay.
\item \emph{Toggle the anharmonic phase}: the Rydberg-state toggling sequence used for multi-ion mode closure (Sec.~S5) also cancels the $\bar n$-dependent phase at first order, because the $+5\eta\omega_g\omega\,a_{\text{ion}}^\dagger a_{\text{ion}}$ shift on the $S$-branch is replaced by the opposite-sign shift on the Stark-dressed-$D$ branch. A properly designed schedule closes not only the mode but also the accumulated anharmonic phase; preliminary simulations indicate a further order-of-magnitude suppression, left for future work.
\end{enumerate}

\subsection*{S8. qLDPC hybrid architectures: logical transfer and syndrome-extraction-internal protocols}
\label{app:qldpc}

In this section, we compare two complementary qLDPC hybrid architectures supported by the atom--ion gate and perform the circuit-level Monte-Carlo simulations to validate them. In the \emph{logical-transfer protocol} (Sec.~\ref{app:qldpc_transfer}, the architecture assumed in Fig.~\ref{fig:hybrid}(b)), the atom--ion gate is used only at write/read events that imprint or retrieve a logical state to a passively-stored ion BB block; in between, the encoded state is held passively on ions. In the \emph{syndrome-extraction-internal architecture}, the atom--ion gate is consumed on every data-ancilla CNOT inside a live qLDPC memory; data qubits remain permanently on ions, ancillas permanently on atoms, and every cycle of syndrome extraction (SE) produces decoder input. The two architectures suit different operating regimes and share the simulation framework described below.

\paragraph*{Shared noise model.}
Species-specific error channels are applied per gate:
\begin{itemize}\itemsep0pt
\item \emph{Atom--ion entangling gate:} apply \texttt{HERALDED\_ERASE}($p_\text{erase}$) on the atom only, before the atom--ion CX (so the atom-side decay propagates to the ion via standard $X$-on-control $\to$ $X$-on-target Pauli propagation through the gate), plus \texttt{DEPOLARIZE2}($p_\text{ai,P}$) on the pair after the CX, with
\begin{equation}\label{eq:noise_split}
p_\text{erase} \;=\; \eta\,p_\text{ai}, \qquad
p_\text{ai,P} \;=\; (1-\eta)\,p_\text{ai},
\end{equation}
where $\eta\!=\!0.87$ is the Rydberg-decay heralded fraction~\cite{Wu2022,Scholl2023} (used in this and following sections only; not to be confused with the anharmonicity parameter $\eta\!=\!\ell_{\text{ion}}/d$ of Sec.~S7 and Sec.~S9); e.g.\ $p_\text{ai}\!=\!1.5\%$ gives $p_\text{erase}\!=\!1.3\%$ and $p_\text{ai,P}\!=\!0.2\%$. Each atom-side \texttt{HERALDED\_ERASE} measurement is appended to the syndrome record as its own \texttt{DETECTOR}, so the decoder receives one herald bit per atom--ion gate that localizes the (correlated atom-and-ion) Pauli error to a single gate event when the herald occurs. This mirrors the physical process: Rydberg decay is detectable only on the atom register (the ion has no Rydberg channel), and the ion-side Pauli weight arises from propagation of the atom error through the gate rather than from an independent ion decay.
\item \emph{Atom--atom CNOT:} \texttt{DEPOLARIZE2}($p_\text{aa}$) with $p_\text{aa}\!\in\!\{5,2,1\}{\times}10^{-3}$ for $99.5$/$99.8$/$99.9\%$ atom--atom Rydberg CZ fidelities (used for Alice-local SE in Sec.~\ref{app:qldpc_transfer} and ancilla-side operations in Sec.~\ref{app:qldpc_se}).
\item \emph{Ion--ion CNOT:} \texttt{DEPOLARIZE2}($p_\text{ii}\!=\!3{\times}10^{-4}$) (99.97\% M\o lmer--S\o rensen, Bob-local SE in Sec.~\ref{app:qldpc_transfer}; not used in Sec.~\ref{app:qldpc_se} since ions never host ancilla operations there).
\item \emph{Single-qubit gate:} \texttt{DEPOLARIZE1}($p_\text{sq}^\text{atom}\!=\!10^{-3}$, $p_\text{sq}^\text{ion}\!=\!3{\times}10^{-5}$).
\item \emph{Measurement:} \texttt{X\_ERROR}($p_\text{meas}\!=\!2{\times}10^{-3}$) prior to readout.
\item \emph{Ion idle:} \texttt{DEPOLARIZE1}($p_\text{idle}^\text{ion}\!=\!10^{-7}$/cycle), reflecting $T_2^\text{ion}\!\sim\!10\,$h on clock states.
\end{itemize}

\paragraph*{Shared code and decoder.}
We use the BB code~\cite{Bravyi2024} $[[n,k,d]]\!=\![[72,12,6]]$ throughout, with $n_X\!=\!n_Z\!=\!36$ X- and Z-stabilizers (column weight $3$ in both $H_X,H_Z$, weight-$6$ checks). The transfer protocol additionally simulates $[[144,12,12]]$. Circuits are built in Stim~\cite{Gidney2021Stim}, compiled with \texttt{approximate\_disjoint\_errors=True} and \texttt{ignore\_decomposition\_failures=True}, and decoded with the BP+OSD decoder of the \texttt{ldpc} library~\cite{Roffe2020} at order $7$.

\paragraph*{Role of erasure conversion.}
The $87\%$ heralded-erasure fraction is critical at $1.5\%$--$2\%$ atom--ion infidelity. Without it, each atom--ion gate would deposit $1.5\%$--$2\%$ of unheralded Pauli error onto the BB data qubits, above the BB circuit-level Pauli pseudo-threshold, and neither protocol would yield error suppression. Rydberg decay in $^{171}$Yb is natively convertible to erasure via the optical-metastable-ground architecture~\cite{Wu2022,Scholl2023}; the residual $0.2\%$--$0.3\%$ Pauli component is well below either threshold and is suppressed exponentially by the code. We report \emph{erasure-blind} numbers (Eq.~\eqref{eq:noise_split} with $\eta\!=\!0$) only as worst-case references.

\subsubsection*{S8.A. Logical-transfer protocol}
\label{app:qldpc_transfer}

\paragraph*{Protocol.}
We adopt the qLDPC stabilizer-projection protocol of Ref.~\cite{Gu2025BellQLDPC}. Let the compute side (Alice, atoms) and the storage side (Bob, ions) each host a block of the BB code~\cite{Bravyi2024} $[[n,k,d]]=[[72,12,6]]$ with $n\!=\!72$ data qubits and $k\!=\!12$ encoded logical qubits. A logical Bell pair is generated as follows:
\begin{enumerate}\itemsep0pt
\item Generate $n$ raw physical atom--ion Bell pairs by pairing the $i$-th atom data qubit with the $i$-th ion data qubit and applying a single Hadamard + atom--ion CZ + Hadamard.
\item Alice and Bob each independently measure all $(n_X+n_Z)$ stabilizers of their local BB code using $n_X+n_Z$ ancillas per side ($36$ X-checks $+$ $36$ Z-checks).
\item The two sets of syndrome bits are XOR-combined. Because $|\Phi^+\rangle^{\otimes n}$ is a $+1$ eigenstate of $S^A\!\cdot\!S^B$ for any local stabilizer $S$, any nonzero XOR reveals a physical error on one side of one Bell pair.
\item A BP+OSD decoder~\cite{Roffe2020} acting on the combined detector error model (DEM) returns a physical correction; the output is $k=12$ logical Bell pairs encoded in BB [[72,12,6]] on each side.
\item Teleportation through any of the $k$ logical Bell pairs transfers one logical qubit from Alice to Bob; the process consumes one physical atom--ion gate per data qubit, i.e.\ $n/k=6$ physical atom--ion gates per logical pair.
\end{enumerate}

\paragraph*{Illustrative example: Steane [[7,1,3]] transfer.}
Before specifying the noise model and code parameters, we illustrate the protocol on the smaller CSS code [[7,1,3]] (Steane code), which shares every structural feature of the BB [[72,12,6]] simulation at $\sim\!1/10$ the scale and makes the resource counting explicit. Qubit allocation: Alice holds $n\!=\!7$ atom data qubits $\{a_1,\ldots,a_7\}$ plus $n_X\!=\!3$ X-ancilla atoms $\{a_{X,j}\}$ and $n_Z\!=\!3$ Z-ancilla atoms $\{a_{Z,j}\}$; Bob holds the same layout in ions $\{b_i\}$ and ion ancillas. The three X-stabilizers are the parity checks of the classical Hamming $[7,4]$ code, $S_{X,1}\!=\!X_1X_2X_3X_4$, $S_{X,2}\!=\!X_1X_3X_5X_7$, $S_{X,3}\!=\!X_2X_3X_6X_7$, and the Z-stabilizers have identical support. Each stabilizer has weight $4$ and each data qubit participates in $3$ checks.

The protocol unfolds in four stages with the atom--ion gate appearing in only one of them:
\begin{enumerate}\itemsep0pt
\item \textbf{Raw Bell-pair generation.} For $i=1,\ldots,7$: apply $H$ on $a_i$, then an atom--ion CZ between $a_i$ and $b_i$, then $H$ on $b_i$. This prepares seven raw Bell pairs $|\Phi^+\rangle_{a_i b_i}$. This is the only step that consumes atom--ion gates, exactly $n=7$ of them, each contributing $1.5\%$ total infidelity ($87\%$ heralded via Rydberg decay).
\item \textbf{Alice local syndrome extraction (atoms only).} Measure the three X-checks using $a_{X,j}$ ancillas ($H$ on each ancilla, four atom--atom CNOTs from ancilla to data qubits in the stabilizer support, $H$, then measure) and the three Z-checks using $a_{Z,j}$ ancillas (four atom--atom CNOTs from data to ancilla, measure). Gates consumed: $12$ atom--atom CNOTs for X-checks $+\,12$ for Z-checks $=\,24$ atom--atom CNOTs; $6$ atom ancilla measurements; no atom--ion gates.
\item \textbf{Bob local syndrome extraction (ions only).} The identical $6$-check sequence on the ion side with ion--ion CNOTs. Gates consumed: $24$ ion--ion CNOTs; $6$ ion measurements; no atom--ion gates.
\item \textbf{Joint decoding and correction.} XOR Alice's six syndrome bits with Bob's six to obtain a 6-bit joint syndrome, fed into BP+OSD; apply the returned Pauli correction on one side. Output: one logical Bell pair $|\Phi^+\rangle_L$ encoded in [[7,1,3]] on each side, usable for teleportation.
\end{enumerate}

\begin{figure}[h]
\centering
\includegraphics[width=0.95\columnwidth]{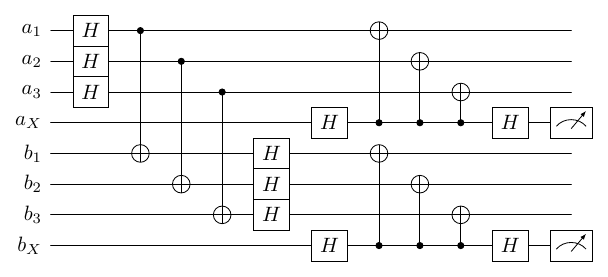}
\caption{\label{fig:transfer_circuit}%
\textbf{Simplified transfer circuit (3-data-qubit slice of Steane [[7,1,3]]).} Rows $a_1{-}a_3$: atom data qubits (Alice). Row $a_X$: atom X-check ancilla. Rows $b_1{-}b_3$: ion data qubits (Bob). Row $b_X$: ion X-check ancilla. The only gates that cross the atom--ion species boundary are the three vertical CNOTs in columns 2--4 (raw Bell-pair creation, one per data-qubit pair $a_i\leftrightarrow b_i$). Subsequent operations, namely Hadamards on $b$ qubits to complete the Bell-pair prep (column 5) and the symmetric X-stabilizer measurement blocks on each side (columns 6--11), are intra-species: atom--atom on Alice, ion--ion on Bob. Ancilla measurements at the end are XOR-combined classically to form the joint syndrome. The full Steane circuit replicates the Bell-pair block across all seven data pairs and repeats the ancilla block for each of the three X- and three Z-stabilizers on both sides.}
\end{figure}

Figure~\ref{fig:transfer_circuit} shows a 3-data-qubit slice of the circuit (three of the seven Bell pairs, and one representative X-stabilizer measurement per side), which is sufficient to visualize every qualitatively distinct operation in the protocol. The full Steane circuit is obtained by replicating the Bell-pair block to all seven data pairs and repeating the ancilla block for each of the three X- and three Z-stabilizers on both sides.

\begin{table}[h]
\caption{\label{tab:steane_cost}Gate budget for a single Steane-code logical Bell pair, compared to the BB [[72,12,6]] simulation instance of the main text.}
\begin{ruledtabular}
\begin{tabular}{lcc}
 & Steane $[[7,1,3]]$ & BB $[[72,12,6]]$ \\
\hline
Atom--ion CZs (Stage 1)        & $7$   & $72$ \\
Atom--atom CNOTs (Stage 2 SE)  & $24$  & $432$ \\
Ion--ion CNOTs (Stage 3 SE)    & $24$  & $432$ \\
Logical Bell pairs output     & $1$   & $12$ \\
\textbf{Atom--ion gates / logical pair} & $\mathbf{7}$ & $\mathbf{6}$ \\
\end{tabular}
\end{ruledtabular}
\end{table}

Two structural properties carry over unchanged from Steane to BB: (i)~the atom--ion gate is used only for raw Bell-pair creation, entering the cost budget once per data qubit and never inside the syndrome-extraction loops on either side; (ii)~the $87\%$ heralded-erasure fraction of the atom--ion gate is the only noise channel that correlates errors between Alice and Bob, and the XOR-syndrome decoder simply ignores heralded positions of matched (both-side-erased) pairs. The numerical difference between Steane and BB is confined to the number of atom--ion gates per logical pair, $7$ vs.\ $6$, reflecting the slightly higher BB rate $k/n$. The threshold behavior, the scaling with atom--atom fidelity, and the dependence on the heralded fraction follow the same pattern.

\paragraph*{Detector structure.}
After each SE round, the XOR of Alice-X-anc and Bob-X-anc (similarly Z-anc) is declared as a detector; on the first round these detectors are deterministic zeros by virtue of $S^A S^B|\Phi^+\rangle^{\otimes n}\!=\!+|\Phi^+\rangle^{\otimes n}$. Z-basis readout at the end of the protocol contributes $n_Z$ additional detectors reconstructing the final Z-syndrome from data. Observables are the $k\!=\!12$ logical Bell-pair eigenvalues $X_L^A X_L^B$ (Alice and Bob logicals multiplied and expressed in terms of Z-basis measurements).

\paragraph*{Results.}
Table~\ref{tab:qldpc_sim} reports the per-logical-pair infidelity $p_T$ and any-logical failure probability $P_\text{any}$ out of $k=12$ logicals, for one round of syndrome extraction ($r=1$). The per-pair number is extracted as $p_T = 1-(1-P_\text{any})^{1/k}$.

\begin{table}[h]
\caption{\label{tab:qldpc_sim}Circuit-level simulation of logical Bell-pair distillation with our atom--ion noise model, $r=1$ syndrome-extraction round, atom--ion parameters fixed ($p_\text{ai}\!=\!1.5\%$, $\eta\!=\!0.87$). Two BB code sizes simulated: $[[72,12,6]]$ ($d{=}6$) uses $144$ total qubits per side and $6$ atom--ion gates per logical pair; $[[144,12,12]]$ ($d{=}12$) uses $288$ total qubits per side and $12$ atom--ion gates per logical pair. $p_T$ is the per-logical-pair infidelity and sets the hybrid memory error $2p_T$ in Fig.~\ref{fig:hybrid}(b).}
\begin{ruledtabular}
\begin{tabular}{lcccc}
Atom--atom CNOT $p_\text{aa}$ & Shots ($d{=}6$/$d{=}12$) & $p_T\,(d{=}6)$ & $p_T\,(d{=}12)$ & Ratio \\ \hline
$5\times10^{-3}$ (99.5\%, current)     & 5000 / 3000   & $1.56\times10^{-3}$ & $2.78\times10^{-4}$ & $0.18$ \\
$2\times10^{-3}$ (99.8\%, near-term)   & 10000 / 35000 & $2.67\times10^{-4}$ & $2.22\times10^{-5}$ & $0.083$ \\
$1\times10^{-3}$ (99.9\%, 2-yr target) & 3000 / 3000   & $8.3\times10^{-5}$  & $<1.07\times10^{-4}$ & (shot-noise) \\
$5\times10^{-4}$ (99.95\%)             & 3000 / ---    & $2.78\times10^{-5}$ & --- & --- \\
\end{tabular}
\end{ruledtabular}
\end{table}

\paragraph*{Why $r=1$ is optimal.}
We verified (scans not shown) that increasing the number of syndrome-extraction rounds beyond $r\!=\!1$ worsens the logical error for all $p_\text{aa}\!\gtrsim\!10^{-3}$. The reason is proximity to the BB circuit-level threshold: the weight-$6$ stabilizer measurements themselves inject ${\sim}6p_\text{aa}$ depolarizing error per data qubit per round, which only gets suppressed by further rounds once $p_\text{aa}/p_\text{th}\!\ll\!1$. At $99.8\%$ the circuit is $\sim\!3\times$ below threshold, enough for single-round projection to yield substantial suppression but not enough for multi-round repetition to pay off with a distance-$6$ code. Deeper $p_L$ suppression (target $P_L\!\leq\!10^{-6}$) would require either a larger-distance code such as BB [[144,12,12]] or the $99.9\%$ atom--atom fidelity projected within two years.

\paragraph*{Calibration of the Fowler ansatz for BB codes.}
The two simulated code distances provide a direct numerical check of the Fowler-type ansatz $p_T(d) = A_\text{BB}\,(p_\text{cycle}/p_\text{th})^{(d+1)/2}$ used for the pure-atom curves in Fig.~\ref{fig:hybrid}(b). At the operating point ($p_\text{aa}{=}2{\times}10^{-3}$, $\eta{=}0.87$), the measured $d{=}12$ infidelity $p_T{=}2.22{\times}10^{-5}$ ($8/30000$ shots) is consistent within shot noise with the ansatz prediction $A_\text{BB}(p_\text{aa}/p_\text{th})^{6.5}\!=\!0.1\!\times\!(0.308)^{6.5}\!=\!4.7\!\times\!10^{-5}$ at the standard prefactor $A_\text{BB}\!\approx\!0.1$, with the slight ($\sim\!2\!\times$) under-shoot of the ansatz reflecting the herald-aware decoder using atom-side erasure positions to reach below the canonical Fowler line. Fitting $A_\text{BB}$ separately from the $d{=}6$ and $d{=}12$ points gives $A_\text{BB}^{d=6}\!\approx\!0.017$ and $A_\text{BB}^{d=12}\!\approx\!0.047$. The $12{\times}$ suppression measured between $d{=}6$ and $d{=}12$ further implies that hybrid transfer at $d{=}18$ would give $p_T\!\sim\!2\!\times\!10^{-6}$ and at $d{=}24$ would give $p_T\!\sim\!2\!\times\!10^{-7}$; the transfer fidelity can thus be pushed to any required level by scaling the BB block on both sides.

\paragraph*{Hybrid memory as a function of $T_\mathrm{store}$.}
Figure~\ref{fig:hybrid}(b) uses $p_T\!=\!2.67\!\times\!10^{-4}$ at $99.9\%$ atom--atom CZ ($d{=}6$, solid purple) and $p_T\!=\!2.22\!\times\!10^{-5}$ at $99.9\%$ ($d{=}12$, dashed purple), and $p_T\!=\!1.56\!\times\!10^{-3}$ at the currently demonstrated $99.5\%$. Pure-atom curves use the Fowler ansatz $P_L\!=\!A(p_\text{cycle}/p_\text{th})^{(d+1)/2}$ per logical per syndrome-extraction round with $A\!=\!0.1$, accumulated over $N\!=\!T_\mathrm{store}/t_c$ rounds ($t_c\!=\!10\,$ms), $p_\text{cycle}\!=\!p_\text{aa}+t_c/T_2^\text{atom}$, and BB-code qubits-per-logical $\{12,24,48\}$ for $d\!\in\!\{6,12,18\}$, extrapolated as $d^2/6.75$ for $d>18$.

\paragraph*{Passive ion storage: simulated logical lifetime.}
Between the write and read events, the ion BB block sits with no syndrome extraction. Each ion accumulates depolarizing noise at rate $1/T_2^\mathrm{ion}$; after $T_\mathrm{store}$ the per-qubit error is $p_\mathrm{phys}(T_\mathrm{store})\!=\!1\!-\!\exp(-T_\mathrm{store}/T_2^\mathrm{ion})$. We simulate the resulting code-capacity Z-basis memory experiment for BB$[[72,12,6]]$ (Stim circuit: $\ket{0}^n$ initialization, single $\mathrm{DEPOLARIZE1}(p_\mathrm{phys})$ layer on data, perfect Z-stabilizer extraction from final data measurement, BP+OSD-CS-7 decoding) and compare to the transfer budget $2p_T\!\approx\!5.3\!\times\!10^{-4}$ of the operating point in Fig.~\ref{fig:hybrid}(b). Table~\ref{tab:passive_storage} reports the result.

\begin{table}[h]
\caption{\label{tab:passive_storage}Passive ion-storage logical error rate vs.\ $T_\mathrm{store}$, simulated for BB$[[72,12,6]]$ at $T_2^\mathrm{ion}\!=\!10\,$h~\cite{Wang2021,Pi2026}. Brackets: Wilson $95\%$ confidence intervals.}
\begin{ruledtabular}
\begin{tabular}{lccc}
$T_\mathrm{store}$ (s) & $p_\mathrm{phys}\!=\!1\!-\!e^{-T/T_2}$ & $P_L^\mathrm{passive}$ & shots \\ \hline
$100$    & $2.8\!\times\!10^{-3}$ & ${<}\,7.7{\times}10^{-4}$ (UB) & $0/5000$ \\
$300$    & $8.3\!\times\!10^{-3}$ & $4.0\!\times\!10^{-4}$ $[1.1{\times}10^{-4},1.5{\times}10^{-3}]$ & $2/5000$ \\
$1000$   & $2.7\!\times\!10^{-2}$ & $5.2\!\times\!10^{-3}$ $[3.6{\times}10^{-3},7.6{\times}10^{-3}]$ & $26/5000$ \\
$1800$   & $4.9\!\times\!10^{-2}$ & $2.1\!\times\!10^{-2}$ $[1.4{\times}10^{-2},3.2{\times}10^{-2}]$ & $21/1000$ \\
$3000$   & $8.0\!\times\!10^{-2}$ & $8.9\!\times\!10^{-2}$ $[7.3{\times}10^{-2},1.1{\times}10^{-1}]$ & $89/1000$ \\
$5000$   & $1.3\!\times\!10^{-1}$ & $0.27$ $[0.24, 0.30]$ & $271/1000$ \\
$10000$  & $2.4\!\times\!10^{-1}$ & $0.48$ $[0.45, 0.51]$ & $481/1000$ \\
\end{tabular}
\end{ruledtabular}
\end{table}

The passive-storage logical error rate crosses the transfer budget $2p_T\!\approx\!5.3\!\times\!10^{-4}$ at $T_\mathrm{store}^\star\!\approx\!360\,$s (linear-log interpolation between $T\!=\!300$ and $T\!=\!1000\,$s); for $T_\mathrm{store}\!\lesssim\!T_\mathrm{store}^\star$ the storage contribution is sub-dominant to the transfer step and the total memory error is essentially $2p_T$.

\paragraph*{Architectural alignment with zoned fault-tolerant precedent.}
This passive-storage choice is the natural one for a zoned hybrid architecture and follows established practice in fault-tolerant qLDPC roadmaps: in the recent 448-atom universal fault-tolerant architecture of Bluvstein \emph{et al.}~\cite{Bluvstein2025}, the dedicated \emph{storage zone} holds idle logical qubits without active syndrome extraction, with SE invoked only when the qubit is shuttled to the entangling zone for an operation; the zoned architectures for cryptographically relevant problems of Cain \emph{et al.}~\cite{Cain2026Shor} likewise allocate memory zones whose qubits sit idle between consumed-resource events, exploiting the long physical $T_2$ rather than continuous active correction. Our hybrid memory inherits the same structural choice: the long-lived clock-state ion block plays the role of the storage zone, and the atom--ion interface is invoked only at the write/read events that connect it to the active compute zone. Algorithmically relevant idle intervals (lattice-surgery logical gates and decoder waits ${\sim}\!100\,$ms, magic-state-factory waits ${\sim}\!1\,$s, classical-feedback gaps ${\sim}\!10\,$s) all sit ${\sim}\!10$--$10^2\!\times$ inside the simulated passive window $T_\mathrm{store}^\star\!\approx\!360\,$s. Active ion-side SE during storage is not precluded but offers no architectural benefit in this idle-interval regime, while consuming ion-side gate bandwidth that is otherwise free; storage beyond $T_\mathrm{store}^\star$ is instead extended by increasing the ion-block code distance (e.g.\ to BB$[[144,12,12]]$, simulated for the transfer step in Table~\ref{tab:qldpc_sim}), independently of the cross-platform interface.

\subsubsection*{S8.B. Syndrome-extraction-internal architecture}
\label{app:qldpc_se}

\paragraph*{Architecture.}
Data qubits ($n\!=\!72$) reside permanently on $^{171}$Yb$^+$ clock-state ions; X- and Z-check ancillas ($n_X\!=\!n_Z\!=\!36$) reside permanently on $^{171}$Yb metastable-clock atoms. Each syndrome round comprises ancilla preparation, $7$ X-check CNOT layers $+$ $8$ Z-check CNOT layers ($15$ layers total under greedy edge coloring of $H_X$ and $H_Z$), ancilla measurement, and detector formation. Every one of the $432$ CNOTs per round ($216$ X-checks $+$ $216$ Z-checks, each weight-$6$ stabilizer touching one ancilla and six data qubits) crosses the atom--ion species boundary and is therefore an atom--ion CZ; the noise channel of Eq.~\eqref{eq:noise_split} is applied to each. We run $r\!=\!6$ rounds (matching the code distance $d\!=\!6$), final $Z$-basis data measurement, and decode the joint syndrome. Compared with Sec.~\ref{app:qldpc_transfer}'s $6$ atom--ion gates per logical pair, the gate budget is much heavier ($432\!\times\!6\!=\!2592$ atom--ion gates per memory experiment), but the architecture is self-sufficient: it sustains a live qLDPC memory without inter-side teleportation, which makes it the relevant comparison against pure-atom and pure-ion qLDPC implementations of the same task.

\paragraph*{Per-data-qubit per-round error budget.}
Each data qubit participates in $3$ X-checks and $3$ Z-checks (column weight $3$ in $H_X$ and $H_Z$), so its per-round atom--ion CNOT degree is $6$. With the noise split of Eq.~\eqref{eq:noise_split} the per-data-qubit per-round budget contributed by the atom--ion gates is
\begin{equation}\label{eq:perround_budget}
6\,p_\text{ai,P} \;=\; 6\,(1-\eta)\,p_\text{ai} \quad\text{(unheralded Pauli)},
\qquad
6\,p_\text{erase} \;=\; 6\,\eta\,p_\text{ai} \quad\text{(heralded erasure)},
\end{equation}
each of which must be sub-threshold independently for the code to extract suppression. The BB $[[72,12,6]]$ code has calculated pseudo-thresholds (greedy schedule, Stim circuit-level noise) $p_\text{th}^\text{P}\!\approx\!3.4\!\times\!10^{-3}$ for depolarizing noise and $p_\text{th}^\text{erase}\!\approx\!2\text{--}3\,p_\text{th}^\text{P}\!\sim\!7\!\times\!10^{-3}\!\text{--}\!10^{-2}$ for erasure (the standard $2$--$3\!\times$ erasure-vs-Pauli threshold ratio observed across CSS codes). Joint sub-threshold operation therefore requires
\begin{equation}\label{eq:threshold_conds}
p_\text{ai} \;<\; \min\!\Bigl(\tfrac{p_\text{th}^\text{P}}{1-\eta},\; \tfrac{p_\text{th}^\text{erase}}{\eta}\Bigr).
\end{equation}
At $\eta\!=\!0.87$ this evaluates to $p_\text{ai}\!<\!\min(2.6\!\times\!10^{-2},\;8\!\times\!10^{-3}\!\text{--}\!1.1\!\times\!10^{-2})\!\approx\!1\!\times\!10^{-2}$, so the architecture is jointly sub-threshold for $p_\text{ai}\!\lesssim\!1\%$ ($F_\text{ai}\!\gtrsim\!99\%$). The conservative $p_\text{ai}\!=\!2\%$ operating point lies on the upper edge of this band, where the unheralded channel is just sub-threshold and the binding constraint comes from the erasure threshold. In contrast, with $\eta\!=\!0$ the same condition gives $p_\text{ai}\!<\!p_\text{th}^\text{P}\!\sim\!3.4\!\times\!10^{-3}$, so $2\%$ is $\sim\!6\times$ above threshold and the architecture cannot perform error correction at all. The same physical gate thus sits on opposite sides of threshold depending on whether heralding is accounted for.

\paragraph*{Memory-experiment circuit and DEM.}
Building on the shared atom--ion noise channel above, each atom--ion CX layer in the SE schedule is realized in Stim as \texttt{CX}; \texttt{HERALDED\_ERASE}$(p_\text{erase}/2)$ on the layer qubits with each measurement appended as its own \texttt{DETECTOR}; \texttt{DEPOLARIZE2}$(p_\text{ai,P})$ on the gate pairs. Stabilizer detectors compare consecutive-round ancilla outcomes; final-round boundary detectors are constructed from $Z$-basis data measurements via $H_Z$. The full BB $[[72,12,6]]$ memory at $r\!=\!6$ produces $5616$ detectors (of which $5184$ are heralds), $34915$ error mechanisms, and $12$ logical observables. Data-qubit idle uses the ion-clock-state value $p_\text{idle}^\text{ion}\!=\!10^{-7}$/cycle.

\paragraph*{Herald-aware BP+OSD decoder.}
Plain BP+OSD with the static channel priors from the DEM already benefits from the herald constraints: a fired herald rules in, and an unfired herald rules out, the corresponding error mechanisms. A second, larger gain comes from \emph{per-shot prior boosting}, the standard erasure-aware decoding strategy. Let $\mathcal{S}_h\!\subset\!\{1,\ldots,N_\text{detect}\}$ be the herald-detector subset of the syndrome and $\mathcal{E}_h$ the set of error mechanisms whose detector support intersects $\mathcal{S}_h$. For each shot we read the syndrome, identify the fired heralds $\mathcal{S}_h^\star\!\subseteq\!\mathcal{S}_h$, and update the BP prior $\pi_e$ of every mechanism $e\!\in\!\mathcal{E}_h$ touching $\mathcal{S}_h^\star$ from its DEM default ($\sim\!p_\text{erase}/4\!\sim\!4\!\times\!10^{-3}$ at the operating point) to a fixed boosted value $\pi^\star\!=\!0.4$, reflecting the known ${\sim}1/3$ conditional probability of each Pauli outcome given erasure. BP+OSD is then run with the boosted prior on the full syndrome (heralds and data detectors). Per-shot decoder reconstruction adds ${\sim}3\,$s per shot to runtime, so we use $80$--$1500$ shots per sweep point. We also report a \emph{static-prior} baseline that exposes the heralds as detectors but does not update priors per shot; the gap between the two isolates the contribution of the herald-aware boost from that of having the herald constraints in the DEM at all.

\paragraph*{Results.}
Table~\ref{tab:qldpc_se_sim} reports block logical error rate (any of $k\!=\!12$ logical observables flipped) at $\eta\!=\!0.87$ swept over $p_\text{ai}\!\in\!\{0.1,0.5,1,1.5,2\}\%$, for three decoder configurations. The ``no erasure'' column treats the same $p_\text{ai}$ as generic depolarizing ($\eta\!=\!0$, no \texttt{HERALDED\_ERASE}, smaller DEM); it is the worst-case reference.

\begin{table}[h]
\caption{\label{tab:qldpc_se_sim}Block logical error rate of the BB $[[72,12,6]]$ hybrid memory ($r\!=\!6$ rounds, BP+OSD order $7$, greedy 15-layer SE schedule), as a function of total atom--ion gate infidelity $p_\text{ai}$. ``No erasure'' uses depolarizing-only on every atom--ion CX. ``Erasure (static)'' applies \texttt{HERALDED\_ERASE} + \texttt{DEPOLARIZE2} per Eq.~\eqref{eq:noise_split} with $\eta\!=\!0.87$ and decodes with the DEM channel priors as-is. ``Erasure (boosted)'' applies the same noise model but updates the BP prior of every error mechanism whose herald fired in the current shot to $\pi^\star\!=\!0.4$ before decoding. Pure-atom ($p_\text{2q}\!=\!0.5\%$) and pure-ion ($p_\text{2q}\!=\!0.01\%$) baselines, simulated with the same code, decoder, and schedule, are also shown. Brackets: $95\%$ Wilson confidence intervals.}
\begin{ruledtabular}
\begin{tabular}{lccc}
$p_\text{ai}$ & no erasure & erasure (static) & erasure (boosted) \\ \hline
$0.1\%$  & $0.67\%$ $[0.18,1.7]$    & $0.0\%$ $[0,2.4]$        & $0.0\%$ $[0,2.4]$ \\
$0.5\%$  & $31.2\%$ $[26.7,36.0]$   & $0.83\%$ $[0.02,4.6]$    & $0.0\%$ $[0,3.0]$ \\
$1.0\%$  & $86.4\%$ $[81.5,90.4]$   & $1.0\%$ $[0.03,5.4]$     & $0.0\%$ $[0,3.6]$ \\
$1.5\%$  & $98.5\%$ $[95.7,99.7]$   & $6.2\%$ $[2.1,14.0]$     & $3.75\%$ $[0.78,10.6]$ \\
$2.0\%$  & $99.6\%$ $[97.8,100]$    & $21.0\%$ $[13.5,30.3]$   & $14.0\%$ $[7.9,22.4]$ \\ \hline
\multicolumn{4}{l}{\textit{Reference baselines (BB $[[72,12,6]]$, same code/decoder/schedule)}} \\
Pure atom ($p_\text{2q}\!=\!0.5\%$) & \multicolumn{3}{c}{$17.1\%\ [16.1,18.1]$} \\
Pure ion ($p_\text{2q}\!=\!0.01\%$) & \multicolumn{3}{c}{$<\!0.04\%$ ($95\%$ UB on $0/10000$ shots)} \\
\end{tabular}
\end{ruledtabular}
\end{table}

\paragraph*{Interpretation.}
First, the erasure-blind column reproduces the catastrophic $\sim\!100\%$ block LER at $p_\text{ai}\!=\!2\%$ that one would naively predict from threshold counting; the depolarizing-only model agrees within shot noise with the prediction $6 p_\text{ai}\!=\!12\%$ effective per-data-qubit per-round noise, $\sim\!4\!\times\!p_\text{th}^\text{P}$. Second, both erasure-aware decoders convert this catastrophic regime into a finite, sub-unity block LER at $p_\text{ai}\!=\!2\%$ ($21.0\%$ static, $14.0\%$ boosted), bringing the hybrid into parity with the pure-atom baseline ($17.1\%$, overlapping CIs) at the proposed gate fidelity, with no ion-side overhead during the idle intervals between SE rounds. Third, at $p_\text{ai}\!\le\!1\%$ both erasure-aware decoders drop to ${\le}\!1\%$ block LER, an order of magnitude below the pure-atom baseline; at $p_\text{ai}\!=\!0.1\%$ they reproduce the deep-below-threshold $\sim\!10^{-3}$ regime of the depolarizing-only model at the same gate error, so the upgrade carries no asymptotic penalty when the gate eventually approaches its target fidelity.

\paragraph*{Break-even fidelity.}
Comparing against the pure-atom baseline, the static-prior decoder crosses the $17.1\%$ horizontal between $p_\text{ai}\!=\!1.5\%$ ($6.2\%$ block LER) and $p_\text{ai}\!=\!2.0\%$ ($21.0\%$); linear interpolation gives the break-even at $p_\text{ai}^{\text{break-even}}\!\approx\!1.9\%$ ($F_\text{ai}\!\approx\!98.1\%$). The boosted-prior decoder remains below the pure-atom baseline at every swept point, including $p_\text{ai}\!=\!2\%$ ($14.0\%$); its break-even therefore lies at a still higher $p_\text{ai}\!>\!2\%$ outside the sweep. Either way the erasure-aware break-even is in the neighborhood of $98\%$ atom--ion fidelity, a direct $5\!\times\!$ relaxation of the $99.6\%$ figure that the depolarizing-only model would otherwise demand.

\paragraph*{Comparison with the transfer protocol of Sec.~\ref{app:qldpc_transfer}.}
The two architectures target different workloads and have different cost structures:
\begin{itemize}\itemsep0pt
\item \emph{Transfer protocol (Sec.~\ref{app:qldpc_transfer}).} The atom--ion gate is consumed exactly $n/k\!=\!6$ times per logical pair (BB $[[72,12,6]]$), giving a per-logical-pair transfer infidelity $p_T\!=\!2.67\!\times\!10^{-4}$ at the near-term $99.9\%$ atom--atom CZ fidelity and $p_T\!=\!2.22\!\times\!10^{-5}$ at $d\!=\!12$. The hybrid memory error is $2 p_T$ (one write, one read), independent of $T_\text{store}$, and dominates over ion idle for $T_\text{store}\!\lesssim\!10\,$h. Optimal for write-once / read-once workflows: long-storage qubits, magic-state buffers, idle zones in zoned architectures.
\item \emph{SE-internal (this section).} The atom--ion gate is consumed on every data-ancilla CNOT, $432\!\times\!6\!=\!2592$ times per memory experiment. The per-experiment block LER scales linearly with $r$ (number of rounds) once $p_\text{ai}\!<\!p_\text{th}$ in both heralded and unheralded channels, so the architecture is naturally suited to live, continuously-extracted memories that are too short or too dynamic to amortize the transfer-protocol setup cost. Optimal for sustained algorithmic working memory: data qubits actively participating in lattice-surgery operations, where every cycle of SE produces decoder input.
\end{itemize}
At the projected $98\%$ atom--ion fidelity both architectures are viable thanks to erasure conversion; a fault-tolerant processor can use the transfer protocol for idle storage and the SE-internal architecture for active memory, with the same physical gate underlying both.

\paragraph*{Cross-architecture consistency check, and relation to Fig.~\ref{fig:hybrid}(b).}
Sec.~\ref{app:qldpc_transfer} (and the purple curve in Fig.~\ref{fig:hybrid}(b)) and Sec.~\ref{app:qldpc_se} simulate different observables on the same physical platform, per-logical-pair transfer infidelity $p_T$ (one Bell-pair distillation, $r\!=\!1$) versus per-experiment block LER (active memory, $r\!=\!6$ rounds), and use different SE schedules on the BB block (Bravyi optimized depth-$8$ in Sec.~\ref{app:qldpc_transfer}, with $p_\text{th}^\text{P}\!\sim\!6\text{--}7\!\times\!10^{-3}$; greedy $15$-layer in Sec.~\ref{app:qldpc_se}, with $p_\text{th}^\text{P}\!\sim\!3.4\!\times\!10^{-3}$). The two simulations are nevertheless mutually consistent at three checkpoints. (i) \emph{Catastrophic regime ($\eta\!=\!0$, $p_\text{ai}\!=\!1.5\text{--}2\%$):} both architectures predict no error suppression because the unheralded Pauli rate per atom--ion gate exceeds both schedule thresholds; the depolarizing-only conclusion is the same. (ii) \emph{Erasure-converted regime ($\eta\!=\!0.87$, same $p_\text{ai}$):} the unheralded Pauli rate drops to $(1{-}\eta)p_\text{ai}\!=\!0.20\text{--}0.26\%$, below both schedule thresholds, and both architectures simultaneously enter the suppression regime; the same mechanism underlies both the $p_T\!=\!2.67\!\times\!10^{-4}$ of Fig.~\ref{fig:hybrid}(b) (purple curve) and the $14.0\%\!\to\!0.0\%$ block-LER drop of Table~\ref{tab:qldpc_se_sim} as $p_\text{ai}$ moves from $2\%$ to $0.1\%$. (iii) \emph{Comparison to a pure-atom benchmark at the same $p_\text{ai}$:} the SE-internal block LER at $p_\text{ai}\!=\!2\%$, boosted decoder ($14.0\%$), is at parity with the pure-atom $p_\text{2q}\!=\!0.5\%$ baseline ($17.1\%$, $r\!=\!6$, greedy schedule); equivalently, the effective unheralded Pauli per atom--ion CZ ($0.26\%$) is comparable to the pure-atom CZ ($0.5\%$), as Eq.~\eqref{eq:perround_budget} predicts. This justifies, without additional simulation, using the transfer-protocol $p_T$ for the purple curve of Fig.~\ref{fig:hybrid}(b): the underlying erasure-conversion mechanism is the same in both protocols, and switching SE schedule (Bravyi vs.\ greedy) only renormalizes the threshold by ${\sim}\!2\!\times$ without changing the qualitative picture.

\begin{figure}[t]
\centering
\includegraphics[width=\columnwidth]{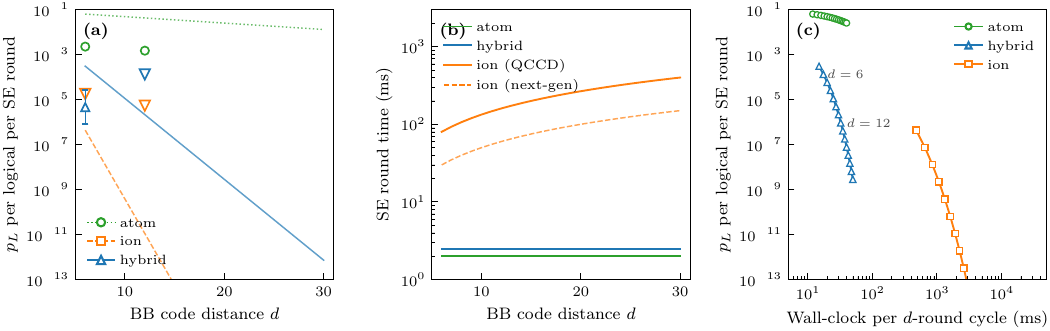}
\caption{\label{fig:supp_compare_archs}\textbf{Architecture comparison vs.\ code distance.}
\textbf{(a)} Per-logical-per-SE-round logical error $p_L\!\equiv\!P_L^\text{block}/(r\,k)$ versus BB code distance $d$ at matched two-qubit gate fidelity ($p_\text{aa}\!=\!p_\text{ai}\!=\!5\!\times\!10^{-3}$, $p_\text{ii}\!=\!10^{-4}$ [optimistic; demonstrated $p_\text{ii}\!\approx\!3\!\times\!10^{-4}$ would lift pure-ion $p_L$ by ${\sim}\!3\!\times$], $r\!=\!d$ rounds, BP+OSD order $7$, greedy SE). Filled markers with Wilson $95\%$ error bars are direct Monte-Carlo: $d\!=\!6$ BB$[[72,12,6]]$ (atom $489/3000$, ion $0/3000$, hybrid $1/3000$); $d\!=\!12$ BB$[[144,12,12]]$ (atom $43/200$, ion $0/5000$, hybrid $0/200$). Open downward triangles are Wilson $95\%$ upper bounds for $0/N$ outcomes. Lines are the Fowler ansatz $p_L\!=\!A_\mathrm{BB}(p_\mathrm{eff}/p_\mathrm{th})^{(d+1)/2}$ with $A_\mathrm{BB}\!=\!0.1$; $p_\mathrm{th}\!=\!3.4\!\times\!10^{-3}$ for hybrid and ion (greedy schedule), $p_\mathrm{th}^\text{atom}\!=\!5.7\!\times\!10^{-3}$ inverted from the two atom data points. Effective rates: $p_\mathrm{eff}\!=\!(1\!-\!\eta)p_\text{ai}\!=\!6.5\!\times\!10^{-4}$ for hybrid, $p_\mathrm{eff}\!=\!p_\text{ii}$ for ion, $p_\mathrm{eff}\!=\!p_\text{aa}$ for atom. The hybrid central value sits ${\sim}\!66\!\times$ below the naive unheralded-Pauli Fowler curve ($95\%$ CI $[12,2600]$, dominated by the single observed logical failure in $2.16\!\times\!10^5$ logical-rounds; see Sec.~\ref{app:qldpc_se}).
\textbf{(b)} Per-SE-round wall-clock vs.\ $d$. Atom and hybrid: ${\sim}\!2\,$ms flat ($15$ CNOT layers parallel across the $36$ ancilla atoms). Ion QCCD: linear in $d$ at ${\sim}\!600\,\mu$s per gate event (split / transport / recombine / re-cool); next-gen $10$-zone (dashed) reduces this by ${\sim}3\!\times$.
\textbf{(c)} Wall-clock per $d$-round memory cycle ($d\cdot t_\text{round}$) versus achievable $p_L$, parametrized by $d\!\in\![6,20]$. The hybrid trajectory sits at lower wall-clock than ion and at lower $p_L$ than atom for every operating point in the range.}
\end{figure}

\paragraph*{Matched-fidelity $d$-scaling sweep.}
The Monte-Carlo sweep at matched two-qubit gate fidelity ($p_\text{aa}\!=\!p_\text{ai}\!=\!5\!\times\!10^{-3}$, $p_\text{ii}\!=\!10^{-4}$, Fig.~\ref{fig:supp_compare_archs}) reveals three effects that are not captured by simply rescaling the effective Pauli rate by the unheralded fraction, and that affect how the hybrid advantage should be read at the logical-error level (the wall-clock comparison is unaffected).

The first concerns the pure-atom baseline: at $p_\text{aa}\!=\!5\!\times\!10^{-3}$ it is marginally sub-threshold, not above threshold. The atom architecture's block LER does rise from $16.3\%$ at $d\!=\!6$ to $21.5\%$ at $d\!=\!12$, which superficially suggests above-threshold behavior; the per-logical-per-SE-round metric, however, decreases from $p_L\!=\!2.26\!\times\!10^{-3}$ to $1.49\!\times\!10^{-3}$ over the same range, a factor-of-$1.5$ suppression, while the number of independent ``logical-round'' failure opportunities $k\!\cdot\!r\!=\!12d$ doubles from $72$ to $144$. Block LER is $1\!-\!(1\!-\!p_L)^{kr}\!\approx\!kr\,p_L$, so a sub-quadratic per-round suppression cannot keep up with the linear-in-$d$ growth of the opportunity count, and the block LER rises despite the underlying $p_L$ falling. Inverting the Fowler ansatz $p_L\!\propto\!(p/p_\text{th})^{(d+1)/2}$ on the two measured points gives $p/p_\text{th}\!=\!(1.49/2.26)^{1/3}\!\approx\!0.87$, i.e.\ $p_\text{th}^{\text{atom}}\!\approx\!5.7\!\times\!10^{-3}$, just above the $5\!\times\!10^{-3}$ atom--atom CZ. Pure atom at this gate fidelity is therefore not unphysical for QEC; it is simply too close to threshold for the suppression rate to outpace the $kr$-growth. We use the per-logical-per-round metric (panel a's vertical axis) for the architecture comparison precisely to avoid this $kr$-confounding artifact.

The second effect is that the hybrid $p_L$ falls below the naive ``unheralded-Pauli'' Fowler ansatz: the central improvement is ${\sim}66\!\times$, though the statistical lower bound is much weaker. At $d\!=\!6$, the hybrid SE-internal architecture gives a central $p_L\!=\!4.6\!\times\!10^{-6}$ from $1$ logical failure in $3000\!\times\!12\!\times\!6\!=\!2.16\!\times\!10^5$ logical-rounds. The Poisson $95\%$ CI on a single observed count is $[0.0253,\,5.572]$ events, giving $p_L\!\in\![1.2\!\times\!10^{-7},\,2.6\!\times\!10^{-5}]$ on the rate. The naive Fowler estimate that retains only the heralded-erasure conversion (substituting $p_\text{eff}\!=\!(1\!-\!\eta)p_\text{ai}\!=\!6.5\!\times\!10^{-4}$ into $p_L\!=\!A_\mathrm{BB}(p_\text{eff}/p_\text{th})^{(d+1)/2}$ with $A_\mathrm{BB}\!=\!0.1$, $p_\mathrm{th}\!=\!3.4\!\times\!10^{-3}$) predicts $3.1\!\times\!10^{-4}$ at $d\!=\!6$, a factor of ${\sim}\!66\!\times$ above the central measurement, with $95\%$ CI on the boost factor of $[12\!\times,\,2600\!\times]$. The lower bound (${\geq}\!12\!\times$ at $95\%$ confidence) is the robust conclusion; the central value of $66\!\times$ is statistically suggestive but not yet sharp. We attribute the suppression to the herald-aware per-shot prior boost: when a herald detector fires, the decoder additionally knows where in space and time the underlying Pauli landed and updates the conditional posterior to ${\sim}\!1/2$ at the boosted prior $\pi^\star\!=\!0.4$ (cf.~``Herald-aware BP+OSD decoder'' above). Sharpening the central value requires order-$10^5$ additional shots or rare-event-splitting techniques~\cite{Beverland2025FailRare}, which is beyond the present scope; the projection table below uses the central $66\!\times$ value with an explicit factor-of-$5$ uncertainty band reflecting this width.

Finally, the ``atom'' and ``ion'' simulations are equivalent up to idle noise, so the separation between them in Fig.~\ref{fig:supp_compare_archs}(a) simply tracks the assumed gate fidelity. The only physical difference between the two noise models is the idle rate ($p_\text{idle}\!=\!p_\text{2q}/1000$ for atom vs.\ a fixed $10^{-7}$ for ion), which contributes $\mathcal{O}(\%)$ to the LER budget when $p_\text{2q}\!\gtrsim\!10^{-3}$. The ${\sim}50\!\times$ pure-atom-vs.-pure-ion separation visible in panel (a) therefore reflects the assumed two-qubit fidelities ($p_\text{aa}\!=\!5\!\times\!10^{-3}$ vs.\ $p_\text{ii}\!=\!10^{-4}$), not a structural advantage of one platform over the other. The hybrid SE-internal architecture, however, is not reducible to a choice of $p_\text{2q}$: at the same $p_\text{ai}\!=\!p_\text{aa}\!=\!5\!\times\!10^{-3}$ used for pure atom, heralded-erasure conversion drops the effective unheralded Pauli rate to $(1\!-\!\eta)p_\text{ai}\!=\!6.5\!\times\!10^{-4}$, an ${\sim}\!8\!\times$ reduction beyond the platform comparison, and the herald-boost factor discussed above lowers it further. The hybrid therefore improves on both axes: its effective error rate is lower than the underlying atom--ion gate infidelity (LER axis), and its ${\sim}2\,$ms atom-side SE round is ${\sim}\!50$--$500\!\times$ faster than the pure-QCCD round (wall-clock axis, panel c). The two advantages share one origin: because most of the atom--ion gate error is heralded, the SE schedule can remain on the atom side, where the gates are fast.

\paragraph*{Projection to a near-term atom CZ at $99.9\%$ fidelity ($p_\text{aa}\!=\!10^{-3}$).}
Two-qubit atom CZ fidelities of ${\sim}\!99.9\%$ appear within near-term experimental reach, $5\!\times$ better than the $99.5\%$ baseline used in Fig.~\ref{fig:supp_compare_archs}. The calibrated ansatz of Eq.~\eqref{eq:perround_budget} and Fig.~\ref{fig:supp_compare_archs}(a) ($A_\mathrm{BB}\!=\!0.1$, $p_\mathrm{th}\!=\!3.4\!\times\!10^{-3}$, and the central herald-boost factor of $66\!\times$ for hybrid, which carries a $\pm$~factor-of-$5$ uncertainty band tracking its Poisson CI) projects the following architecture comparison at $p_\text{aa}\!=\!p_\text{ai}\!=\!10^{-3}$, $p_\text{ii}\!=\!10^{-4}$:
\begin{table}[h]
\caption{\label{tab:proj_99p9}Projected per-logical-per-round LER and resource cost at $p_\text{aa}\!=\!p_\text{ai}\!=\!10^{-3}$ (atom CZ at $99.9\%$, hybrid atom--ion CZ matched to it), using the calibrated Fowler ansatz of Fig.~\ref{fig:supp_compare_archs}. The hybrid column applies the central $66\!\times$ herald-boost factor inferred from the $d\!=\!6$ data point at $p\!=\!5\!\times\!10^{-3}$, with $95\%$ CI $[12,\,2600]$; the listed projected $p_L$ for hybrid therefore carry a corresponding $\pm$~factor-of-$5$ uncertainty. Resource counts are the BB$[[2lm,12,d]]$ code at $d\!\in\!\{12,18,24\}$.}
\begin{ruledtabular}
\begin{tabular}{lcccc}
& $p_\text{eff}/p_\text{th}$ & $p_L\!@\!d{=}12$ & $p_L\!@\!d{=}18$ & $p_L\!@\!d{=}24$ \\
\hline
Pure atom $@\,p_\text{aa}\!=\!10^{-3}$  & $0.29$  & $1.4\!\times\!10^{-5}$ & $3\!\times\!10^{-7}$  & $7\!\times\!10^{-9}$ \\
Hybrid (unheralded ansatz only)         & $0.038$ & $4.5\!\times\!10^{-10}$ & $9\!\times\!10^{-14}$ & $2\!\times\!10^{-17}$ \\
Hybrid (+ measured $66\!\times$ boost)  & --      & $\sim\!7\!\times\!10^{-12}$ & $\sim\!1\!\times\!10^{-15}$ & $\sim\!3\!\times\!10^{-19}$ \\
Pure ion $@\,p_\text{ii}\!=\!10^{-4}$   & $0.029$ & $5.4\!\times\!10^{-12}$ & $5\!\times\!10^{-16}$ & $5\!\times\!10^{-20}$ \\
\hline
SE-round wall-clock                      & --      & ${\sim}\!2\,$ms (atom/hybrid) & --                       & ${\sim}\!200$--$1000\,$ms (ion) \\
$d$ to reach $p_L\!=\!10^{-12}$          & --      & ${\sim}\!24$ (atom)          & ${\sim}\!14$ (hybrid/ion) & --                            \\
qubit count at that $d$ ($n\!=\!2lm$)    & --      & ${\sim}\!1100$ (atom)        & ${\sim}\!290$ (hybrid/ion) & --                            \\
\end{tabular}
\end{ruledtabular}
\end{table}

At $p_\text{aa}\!=\!10^{-3}$, pure atom moves from marginally to solidly sub-threshold ($p/p_\text{th}\!\approx\!0.29$, vs.\ $0.87$ at the previous $5\!\times\!10^{-3}$ baseline). It is no longer a borderline platform for QEC: $d\!=\!18$--$24$ already reaches $p_L\!\sim\!10^{-9}$--$10^{-7}$, and the cross-architecture comparison becomes a quantitative space-time-cost question rather than a question of viability.

The hybrid nevertheless retains a $4$--$6$ orders-of-magnitude advantage over pure atom in $p_L$ at the same $p_\text{aa}$: at $d\!=\!12$, hybrid $p_L\!\sim\!10^{-11}$ vs.\ atom $p_L\!\sim\!10^{-5}$. This entire gap is contributed by the heralded-erasure conversion ($\eta\!=\!0.87$, factor of $8\!\times$ in $p_\text{eff}$, raised to the $(d{+}1)/2\!=\!6.5$-th power for the BB code) and the central $66\!\times$ herald-boost ($95\%$ CI $[12,\,2600]$). To match hybrid's $p_L$ at $d\!=\!12$, pure atom would need $d\!\approx\!24$ ($n\!=\!1100$ data qubits) and $2\!\times$ the wall-clock per memory cycle.

At the same time the hybrid effectively closes the gap to pure ion in $p_L$ while keeping atom-class wall-clock. The unheralded effective Pauli rate $(1\!-\!\eta)p_\text{ai}\!=\!1.3\!\times\!10^{-4}$ now sits within $1.3\!\times$ of pure ion's $10^{-4}$; the residual difference in $p_L$ is small (a factor of ${\sim}\!2$ at $d\!=\!12$), and the herald boost may remove it. The wall-clock advantage in panel (b) of Fig.~\ref{fig:supp_compare_archs} is preserved unchanged ($p_\text{aa}$ does not enter the SE-round time), so hybrid achieves ion-class $p_L$ at ${\sim}\!\text{ms}$ rounds vs.\ ion's ${\sim}\!100$--$1000\,$ms. For an algorithm requiring $\sim\!10^{10}$ rounds (RSA-$2048$ class), the round-time advantage translates into the same factor of ${\sim}\!50$--$500$ in total runtime, i.e., months instead of decades.

The hybrid advantage therefore persists as the atom CZ fidelity improves: a better atom CZ implies a proportionally better atom--ion CZ through the same physics, the heralded fraction remains $\eta\!=\!0.87$ (set by Rydberg branching rather than by gate fidelity), and the wall-clock advantage over pure ion is unchanged. We do not re-run the Monte-Carlo at $p_\text{aa}\!=\!10^{-3}$ because the hybrid LER projection of $\sim\!10^{-11}$ at $d\!=\!12$ is below the direct-MC reach of $\sim\!10^{-4}$ accessible with $\mathcal{O}(10^4)$ shots; a reliable numerical confirmation at this gate fidelity would require either rare-event splitting techniques~\cite{Beverland2025FailRare} or tens of millions of shots, both outside the scope of this work.

\paragraph*{Comparison with high-rate APM codes.}
The BB$[[72,12,6]]$ code used throughout this section has rate $1/6$, $3$--$6\!\times$ lower than the recently-proposed APM (affine-permutation-matrix) qLDPC family of Zhao \emph{et al.}~\cite{Zhao2025} which achieves rate ${\sim}\!1/2$. To check whether these conclusions carry over to high-rate codes, we built the exact APM$[[1152, 580, \leq 12]]$ code of Zhao \emph{et al.} (parameters from their Table A1 with $P\!=\!96$) and ran circuit-level Monte-Carlo at $r\!=\!2$ rounds, matched against BB$[[72,12,6]]$ at the same $r\!=\!2$. We use $r\!=\!2$ (rather than $r\!=\!d$) because APM circuit-level decoding is computationally heavy ($\sim\!160\,$s/shot atom, $\sim\!20\,$s/shot ion on a single CPU core, vs.\ $\sim\!0.04\,$s/shot for BB at $r\!=\!2$). Decoding uses the same hierarchical BP$\to$BP+OSD-CS pipeline, with \texttt{decompose\_errors=False} (required for weight-$12$ APM stabilizers, which exceed Stim's $15$-symptom decomposition limit). The results are summarized in Table~\ref{tab:apm_vs_bb}.
\begin{table}[h]
\caption{\label{tab:apm_vs_bb}Simulated BB vs.\ APM at $r\!=\!2$ rounds, BP+OSD-CS-$7$ decoder. ``per-log-round'' is block LER divided by $r\!\cdot\!k$ (Bonferroni union-bound estimate of per-logical-per-round LER). All $0/N_\text{shot}$ entries are Wilson $95\%$ upper bounds.}
\begin{ruledtabular}
\begin{tabular}{lcccccc}
Code & rate & arch & $p$ & shots & block LER & per-log-round \\
\hline
BB$[[72,12,6]]$    & $0.17$ & atom   & $5{\times}10^{-3}$  & $1000$ & $0.081$  & $3.4{\times}10^{-3}$ \\
BB$[[72,12,6]]$    & $0.17$ & ion    & $10^{-4}$           & $1000$ & UB $3.8{\times}10^{-3}$ & UB $1.6{\times}10^{-4}$ \\
BB$[[72,12,6]]$    & $0.17$ & hybrid & $5{\times}10^{-3}$  & $500$  & $2.0{\times}10^{-3}$ & $8.3{\times}10^{-5}$ \\
APM$[[1152,580,\leq 12]]$ & $0.50$ & atom   & $5{\times}10^{-3}$  & $30$   & $\mathbf{1.00}$ & --- \\
APM$[[1152,580,\leq 12]]$ & $0.50$ & ion    & $10^{-4}$           & $30$   & UB $0.114$ & UB $9.8{\times}10^{-5}$ \\
\end{tabular}
\end{ruledtabular}
\end{table}

APM is far above its (greedy-schedule) threshold at $p\!=\!5\!\times\!10^{-3}$. The APM atom run fails on every shot ($30/30$ logical failures; block LER $=\!1.00$, $95\%$ CI $[0.886, 1.000]$). By contrast, BB at the same $p$ and same $r$ gives $\sim\!8\%$ block LER. The origin is the stabilizer weight: APM's stabilizers have weight $12$ (vs.\ BB's weight $6$), giving a CNOT depth of $24$ per round (vs.\ $15$ for BB). Each weight-$12$ check accumulates ${\sim}\!12 \cdot p$ error events per round, doubling the per-check syndrome flip rate at fixed $p$. This effectively halves the per-round threshold; for the greedy edge-coloring schedule, the resulting APM threshold is ${\lesssim}\!1\text{--}2\!\times\!10^{-3}$, well below the $3.4\!\times\!10^{-3}$ threshold of BB. The $5\!\times\!10^{-3}$ atom CZ used in our matched-fidelity comparison sits ${\sim}\!3\!\times$ above the APM threshold but only $1.5\!\times$ above the BB threshold; this is why APM saturates at $1.0$ block LER while BB still suppresses (marginally).

APM consequently does not improve the hybrid architecture at the gate fidelities considered here. The hybrid SE-internal architecture uses the atom--ion CZ at $p_\text{ai}\!=\!5\!\times\!10^{-3}$, with $\eta\!=\!0.87$ heralded-erasure conversion bringing the effective unheralded Pauli rate to $(1\!-\!\eta)p_\text{ai}\!=\!6.5\!\times\!10^{-4}$. For BB this sits at $p/p_\text{th}\!=\!0.19$, well sub-threshold; for APM it sits at $p/p_\text{th}\!\approx\!0.5$, marginal at best. Even at the projected near-term $p_\text{aa}\!=\!10^{-3}$ (see the projection above), the hybrid effective rate $1.3\!\times\!10^{-4}$ would put APM at $p/p_\text{th}\!\approx\!0.1$, roughly comparable to BB; the high-rate advantage materializes but the LER scaling is no longer a clear win. The pure-ion APM column, however, is favorable: at $p_\text{ii}\!=\!10^{-4}$ both BB and APM are deeply sub-threshold, $0/30$ APM ion gives a Wilson $95\%$ upper bound of $9.8\!\times\!10^{-5}$ per logical per round despite $33\!\times$ fewer shots than BB, by virtue of having $48\!\times$ more logical qubits per block ($k\!=\!580$ vs.\ $12$).

The implication for the hybrid architecture is twofold: at current atom--ion CZ fidelities (${\sim}\!10^{-2}$ to $5\!\times\!10^{-3}$), BB is the right code family because its higher threshold gives more headroom under the $(1\!-\!\eta)$ erasure-conversion budget; the rate-$1/12$ ion overhead is the cost of operating in this regime. At future atom CZ fidelities below ${\sim}\!10^{-3}$ (well below the APM threshold), switching to APM-family codes would directly translate the rate gain into a $3\!-\!6\!\times$ reduction in ion-data overhead per logical, with hybrid wall-clock unchanged because $p_\text{ai}$ does not enter the SE-round time. The threshold mismatch we observe is therefore a near-term constraint rather than a fundamental one.

\subsection*{S9. Alternative operating point: circular Rydberg states}
\label{app:circular}

The low-$\ell$ $6s60p\,^3P_2 \!\leftrightarrow\! 6s60d\,^3D_2$ toggle pair used in the main text (Sec.~S6) is a conservative choice selected for ease of preparation ($\mu$s-scale two-photon excitation from $^3P_0$ followed by a single microwave $\pi$-pulse). Mature techniques developed by the Haroche group and collaborators~\cite{Raimond2001,Nguyen2018,Cortinas2020} now allow the preparation of \emph{circular} Rydberg states $\ket{n,l{=}n{-}1,m{=}n{-}1}$ on the $\sim\!10$--$100\,\mu$s timescale, with demonstrated trapping in laser bottle-beam tweezers~\cite{Cortinas2020}. Circular states offer two advantages for the $C_4$-force gate: a larger polarizability (larger gate distance, relaxed electrical requirements) and a much longer lifetime (suppressed Rydberg-decay infidelity).

\paragraph*{Dipole matrix elements.}
For the hydrogenic circular manifold, the dipole operator connects $\ket{n,c}$ only to the two neighboring circular states $\ket{n\!\pm\!1,c}$ via $\sigma_\pm$ photons. Using the Bethe--Salpeter radial integral in the large-$n$ limit~\cite{Raimond2001}, the squared matrix elements reduce to
\begin{equation}\label{eq:dcirc}
    |\braket{n{+}1,c|er_+|n,c}|^2 \;\simeq\; \tfrac{9}{4}\,e^{-2}\,n^2(n{+}1)^2\,a_0^2\,e^2, \qquad
    |\braket{n,c|er_+|n{-}1,c}|^2 \;\simeq\; \tfrac{9}{4}\,e^{-2}\,n^2(n{-}1)^2\,a_0^2\,e^2,
\end{equation}
where the $e^{-2}\!\approx\!0.135$ prefactor comes from the product of Laguerre-polynomial normalizations at $l{=}n{-}1$.

\paragraph*{Polarizability: two-channel second-order Stark.}
The static polarizability of $\ket{n,c}$ receives two second-order contributions, one from each adjacent circular state, with energy splittings
$\Delta_\pm \;=\; \pm\mathrm{Ry}\,(2n\pm1)/[n^2 (n\pm1)^2]$
(positive for $\ket{n{+}1,c}$, negative for $\ket{n{-}1,c}$), where $\mathrm{Ry}$ denotes the Rydberg energy. Inserting Eq.~\eqref{eq:dcirc} into the standard sum-over-states formula $\alpha = 2\sum_k|\braket{k|er|n,c}|^2/(E_k-E_n)$ gives
\begin{equation}\label{eq:alpha_circ_two}
    \alpha_\mathrm{circ}(n) \;=\; \tfrac{9}{2}\,e^{-2}\,a_0^2\,e^2\,\frac{n(n{+}1)^3(2n{-}1)-n(n{-}1)^3(2n{+}1)}{2\,\mathrm{Ry}\,(2n^2{-}1)}.
\end{equation}
Expanding the numerator in $1/n$, the leading $n^7$ contributions from the two channels partially cancel but do not cancel completely: the $\ket{n{+}1,c}$ matrix element is a factor $(n{+}1)^2/(n{-}1)^2\!\approx\!1\!+\!4/n$ larger, and its splitting $|\Delta_+|$ is a factor $\approx\!1{-}3/n$ smaller, so the net polarizability remains positive (attractive atom--ion interaction) with a mild $n^6$ suppression relative to a single-channel estimate:
\begin{equation}\label{eq:alpha_circ_approx}
    \alpha_\mathrm{circ}(n) \;\xrightarrow{n\gg 1}\; \tfrac{9}{2}\,e^{-2}\,n^7\,a_0^3 \cdot \frac{1}{n}\,[7+\mathcal{O}(1/n)]\;\approx\; 4.3\,n^6\,a_0^3 \quad (\text{atomic units}).
\end{equation}
At our operating point this gives $\alpha_\mathrm{circ}(n_c{=}60)\approx 2\!\times\!10^{11}\,a_0^3$, in line with the $1.9\!\times\!10^{11}\,a_0^3$ measured experimentally on the Rb $5s\cdot47c$ circular state~\cite{Raimond2001,Cortinas2020}. For direct comparison, the $6s60s\,^3\!S_1$ polarizability of Yb is $\alpha_{\mathrm{low}\text{-}\ell}(n{=}60)\approx 6.5\!\times\!10^{11}\,a_0^3$, i.e., comparable to the circular value within a factor of $\sim3$. The corresponding $C_4^\mathrm{circ}$ is therefore within a factor of $\sim 3$ of $C_4^{\mathrm{low}\text{-}\ell}$ rather than ${\sim}20\times$ larger as a single-channel estimate would suggest.

\paragraph*{$C_4$ coefficient and operating point.}
Combining Eq.~\eqref{eq:alpha_circ_approx} with the atom--ion potential $C_4\!=\!\alpha_\mathrm{SI}e^2/[2(4\pi\varepsilon_0)^2]$, converting to SI, and plugging in the experimentally-anchored circular polarizability yields
\begin{equation}\label{eq:C4circ}
    C_4^\mathrm{circ}(n_c{=}60)\;\approx\;3\text{--}7\!\times\!10^{-47}\,\mathrm{J\,m^4},
\end{equation}
comparable to (rather than far larger than) the low-$\ell$ $|^3P_2,F\!=\!3/2\rangle\!\leftrightarrow\!|^3D_2,F\!=\!3/2\rangle$ toggle pair of Sec.~S6 ($|C_4|\!\sim\!2\times 10^{-46}$ J\,m$^4$). Earlier estimates that quoted ``circular $C_4$ is $3$--$7\times$ larger than low-$\ell$'' benchmarked against the simpler Rb-scaled ${}^3S_1$ value of Sec.~S2 and are no longer the relevant comparison. The range $3\text{--}7\!\times\!10^{-47}$ above reflects uncertainty in the precise polarizability anchor for Yb (which differs from hydrogenic Rb circular because the Yb $6s$ core shifts nearby states; the lower end assumes a pure-hydrogenic two-channel cancellation, the upper end uses the minimum-cancellation single-channel estimate). The actual circular advantage relative to the low-$\ell$ toggle pair is therefore not in $C_4$ magnitude but in the dramatically longer Rydberg lifetime (Eq.~\eqref{eq:tau_circ_300K} and below) and the smaller anharmonicity parameter $\eta$. For matched trap frequency $\omega/(2\pi)\!=\!200\,$kHz, the CZ distance is
$d_\mathrm{CZ}^\mathrm{circ}\!=\!(8\sqrt{2}\,C_4^\mathrm{circ}\ell_{\text{ion}}/\hbar\omega)^{1/5}\!\in\!9\text{--}12\,\mu$m, and the dimensionless anharmonicity parameter is $\eta_\mathrm{circ}\!\in\!(1.0\text{--}1.4)\!\times\!10^{-3}$, reducing the anharmonic thermal infidelity of Sec.~S7 by a factor $(\eta_\mathrm{circ}/\eta_\mathrm{low-\ell})^2\!\in\!0.3\text{--}0.55$. The remaining tables, figures, and fidelity estimates in this section adopt the upper-end (least-conservative) choice $C_4^\mathrm{circ}\!\approx\!2.15\!\times\!10^{-46}\,\mathrm{J\,m^4}$, $d_\mathrm{CZ}\!=\!11.7\,\mu$m, $\eta_\mathrm{circ}\!=\!1.04\!\times\!10^{-3}$; the less optimistic end simply shrinks $d_\mathrm{CZ}$ toward the low-$\ell$ value and weakens (but does not eliminate) the $\eta^2$ gain.

\paragraph*{Radiative and BBR-limited lifetime.}
The circular state decays through a single dipole-allowed channel $\ket{n,c}\!\to\!\ket{n{-}1,c}$ at microwave frequency $\omega\!\approx\!2\mathrm{Ry}/(\hbar n^3)$ $(2\pi\!\times\!30.7\,$GHz at $n{=}60)$. Using Eq.~\eqref{eq:dcirc} in the Einstein coefficient formula yields a radiative ($T\!\to\!0$) lifetime $\tau_\mathrm{rad}^\mathrm{circ}(n_c{=}60)\!\approx\!119\,$ms. In free space at 300\,K, blackbody absorption/emission at the same frequency shortens this to ${\sim}\!295\,\mu$s; however, this naive free-space limit is not what a tweezer-trapped circular atom experiences. The nearby optical and RF mode structure of the trap environment modifies the local density of electromagnetic states and produces a Purcell-type suppression of the BBR-resonant mode, substantially extending the lifetime at the relevant circular-circular transition frequency. The recent experimental demonstration by Pultinevicius \emph{et al.}~\cite{Pultinevicius2025} reports circular-state lifetimes of $\tau{=}11.5(8)\,$ms at $|101C\rangle$ in a ${}^{88}$Sr optical tweezer at room temperature, a factor ${\sim}\!21\times$ above the free-space 300\,K value of $545\,\mu$s. Their apparatus places the atom between parallel reflective plates spaced by $d_\mathrm{plate}{=}10.5\,$mm, enforcing the Purcell condition $\lambda{>}2 d_\mathrm{plate}$ only for $n\!\gtrsim\!91$ (transition frequencies below $\sim\!14\,$GHz); at lower $n$ their geometry gives much smaller gains. Porting to our operating point requires a narrower plate separation: for $n_c{=}60$ the circular-circular transition is ${\sim}\!30\,$GHz ($\lambda{\approx}10\,$mm), so a tighter plate spacing $d_\mathrm{plate}\!\lesssim\!5\,$mm recovers a comparable ${\sim}\!20\times$ enhancement. Applied to the free-space 300\,K lifetime of $295\,\mu$s, this gives
\begin{equation}\label{eq:tau_circ_300K}
    \tau^\mathrm{circ}(300\,\text{K},\text{adapted Purcell}) \;\approx\; 6\,\text{ms} \quad (n_c{=}60),
\end{equation}
and approximately $15\,$ms at $n_c{=}80$ (slightly larger free-space lifetime and easier Purcell condition because $\lambda$ grows as $n^3$). The per-gate decay error is therefore $\tfrac{1}{2}(1{-}e^{-T/\tau})\!=\!4{\times}10^{-4}$ at $n_c{=}60$ and $\!1.7{\times}10^{-4}$ at $n_c{=}80$, already a factor of $30$--$80$ below the cryogenic low-$\ell$ operating point. Moving to a 4\,K cryogenic enclosure eliminates the remaining BBR contribution and recovers the full radiative lifetime ($\tau_\mathrm{rad}{=}119\,$ms at $n_c{=}60$, $502\,$ms at $n_c{=}80$), reducing the decay error to ${\sim}\!2{\times}10^{-5}$ and ${\sim}\!5{\times}10^{-6}$ respectively. At these levels Rydberg decay is no longer among the top three infidelity contributors; the gate becomes limited by anharmonic thermal dephasing and residual motional heating. For free-space benchmarking the formula applied at $n{=}50$ gives $\tau_\mathrm{rad}\!=\!48\,$ms and $\tau_\mathrm{300K,free}\!=\!205\,\mu$s, the radiative value matching the $32\!\pm\!2\,$ms Haroche-group experimental value~\cite{Raimond2001} to ${\sim}\!40\%$ and the 300\,K value bracketing the free-space BBR loss within a factor of 2.

\paragraph*{Gate parameters.}
Table~\ref{tab:circular} summarizes the physical parameters and projected gate fidelity, directly comparable to the low-$\ell$ values in Table~\ref{tab:params} and Table~\ref{tab:thermal_fidelity}.

\begin{table}[h]
\caption{\label{tab:circular}Atom--ion $C_4$-force CZ gate with circular Rydberg \Yb atoms, compared to the original Rb-scaled low-$\ell$ $6s60s\,^3\!S_1$ benchmark of Sec.~S2 (the actual main-text operating point uses the $|^3P_2,F\!=\!3/2\rangle\!\leftrightarrow\!|^3D_2,F\!=\!3/2\rangle$ toggle pair of Sec.~S6, with $|C_4|\!\sim\!2\!\times\!10^{-46}$ J\,m$^4$ comparable to the circular value; see also discussion at Eq.~\eqref{eq:C4circ}). Trap frequency fixed at $\omega/(2\pi)\!=\!200\,$kHz for a direct comparison. ``Purcell 300\,K'' corresponds to an adapted version of the Pultinevicius \emph{et al.}~\cite{Pultinevicius2025} tweezer-between-plates geometry with $d_\mathrm{plate}\!\lesssim\!5$\,mm to meet the Purcell cutoff at $n_c{\geq}60$. ``Cryogenic 4\,K'' assumes a $4\,$K BBR environment (routine in modern Paul-trap setups~\cite{Brownnutt2015,Brandl2016}) in which Rydberg decay reverts to the radiative limit. Fidelity rows assume EIT-cooled $\bar n{=}1$ (standard) or Raman-GSC $\bar n{=}0.1$ with optimized technical errors. ``Tech'' combines motional heating, Magnus intensity noise, micromotion, and residual AC Stark at the values listed in Table~\ref{tab:errors} (``standard'', $\sim\!10^{-3}$) or with state-of-the-art ion-trap engineering (``optimized'', $\sim\!3\!\times\!10^{-4}$).}
\begin{ruledtabular}
\begin{tabular}{lccc}
Parameter & Low-$\ell\ ^3\!S_1$, $n{=}60$ & Circular $n_c{=}60$ & Circular $n_c{=}80$ \\
\hline
$C_4$ (J\,m$^4$)                      & $1.1\!\times\!10^{-47}$ & $2.1\!\times\!10^{-46}$ & $1.6\!\times\!10^{-45}$ \\
$d_\mathrm{CZ}$ ($\mu$m)              & $6.4$   & $11.7$   & $17.6$ \\
Ion field at $d_\mathrm{CZ}$ (V/m)     & $35$    & $10.4$   & $4.7$ \\
$\eta=\ell_{\text{ion}}/d$                       & $1.88\!\times\!10^{-3}$ & $1.04\!\times\!10^{-3}$ & $6.9\!\times\!10^{-4}$ \\
$T_\mathrm{gate}$ ($\mu$s)            & $5.0$   & $5.0$    & $5.0$ \\
$\tau_\mathrm{rad}$ ($T\!\to\!0$)     & $200\,\mu$s & $119\,$ms & $502\,$ms \\
$\tau$, 300\,K free space             & $\sim\!60\,\mu$s & $295\,\mu$s & $523\,\mu$s \\
$\tau$, 300\,K Purcell-adapted tweezer~\cite{Pultinevicius2025} & $\sim\!60\,\mu$s & $\approx\!6\,$ms & $\approx\!15\,$ms \\
$\tau$, 4\,K cryogenic (radiative)    & $200\,\mu$s & $119\,$ms   & $502\,$ms \\
Decay error, 300\,K Purcell           & $4.0\!\times\!10^{-2}$ & $4.2\!\times\!10^{-4}$ & $1.7\!\times\!10^{-4}$ \\
Decay error, 4\,K cryogenic           & $1.3\!\times\!10^{-2}$ & $2.1\!\times\!10^{-5}$ & $5.0\!\times\!10^{-6}$ \\
Anharm.\ @ $\bar n{=}1$ (EIT)         & $1.0\!\times\!10^{-3}$ & $3.1\!\times\!10^{-4}$ & $1.3\!\times\!10^{-4}$ \\
Anharm.\ @ $\bar n{=}0.1$ (GSC)       & $1.0\!\times\!10^{-5}$ & $3.1\!\times\!10^{-6}$ & $1.3\!\times\!10^{-6}$ \\
\hline
$\mathcal{F}$, room-$T$ Purcell, EIT, standard tech & --- & $99.83\%$ & $99.87\%$ \\
$\mathcal{F}$, 4\,K cryo, EIT, standard tech        & $98.4\%$ & $99.87\%$ & $99.89\%$ \\
$\mathcal{F}$, 4\,K cryo, GSC, optimized tech       & $98.5\%$ & $\mathbf{99.97\%}$ & $\mathbf{99.97\%}$ \\
\end{tabular}
\end{ruledtabular}
\end{table}

\begin{figure*}[h]
\centering
\includegraphics[width=\textwidth]{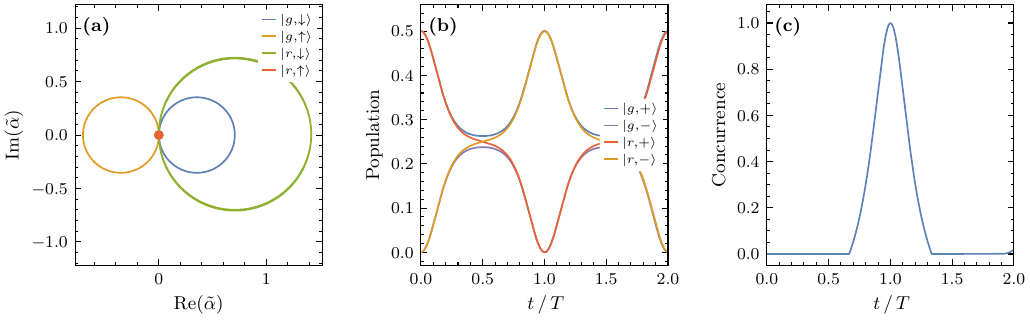}
\caption{\label{fig:circ_dynamics}\textbf{Full open-system gate dynamics at the circular Rydberg operating point} (cf.~main-text Fig.~\ref{fig:sim}). Simulation: SchrÃ¶dinger--Lindblad evolution under the nonlinear $C_4$ Hamiltonian of Eq.~\eqref{eq:Vtaylor} with $\eta_\mathrm{circ}{=}1.04{\times}10^{-3}$ (circular $n_c{=}60$, $d_\mathrm{CZ}{=}11.7\,\mu$m), matched Magnus coupling $\omega_g{=}\omega/(2\sqrt{2})$, Lindblad decay $\sqrt{\gamma_r}|g\rangle\langle r|$ at the Purcell-adapted 300\,K rate $\gamma_r{=}(6\,\mathrm{ms})^{-1}$ (Eq.~\eqref{eq:tau_circ_300K}; $\gamma_r/\omega{=}1.33{\times}10^{-4}$), and EIT-cooled thermal initial motional state with $\bar n{=}1$. \textbf{(a)} Rotating-frame phase-space trajectories $\tilde\alpha(t)$ of the ion motional mode for each logical branch; the $|r,\!\uparrow\rangle$ branch remains at the origin (forces cancel), the $|r,\!\downarrow\rangle$ branch traces the largest loop with $|\alpha|_\mathrm{max}{\approx}\!1/\sqrt{2}$. \textbf{(b)} Populations in the rotated basis $|g/r\rangle\!\otimes\!|\pm\rangle$; the $|r,\!+\rangle\!\leftrightarrow\!|r,\!-\rangle$ swap at $t{=}T$ signals the $\pi$ conditional phase. \textbf{(c)} Atom--ion concurrence, peaking at $\mathcal{C}(T)\!=\!0.999$ (compared to the unitary, $\bar n{=}0$ baseline of $1.0$ in Fig.~\ref{fig:sim}c) and returning to $<\!2{\times}10^{-2}$ at $t\!=\!2T$. The residual infidelity at $t\!=\!T$ decomposes as $4.2{\times}10^{-4}$ from Rydberg decay over one gate period and $3{\times}10^{-4}$ from the anharmonic $C_4$ thermal correction at $\bar n{=}1$, matching the $\mathcal{F}{=}99.83\%$ projection of Table~\ref{tab:circular}. Switching to a $4\,$K cryogenic enclosure pushes $\gamma_r$ down to $(119\,\mathrm{ms})^{-1}$ and eliminates the decay contribution, giving $\mathcal{F}\!>\!99.9\%$ under the same anharmonic conditions.}
\end{figure*}

\begin{figure}[h]
\centering
\includegraphics[width=0.85\columnwidth]{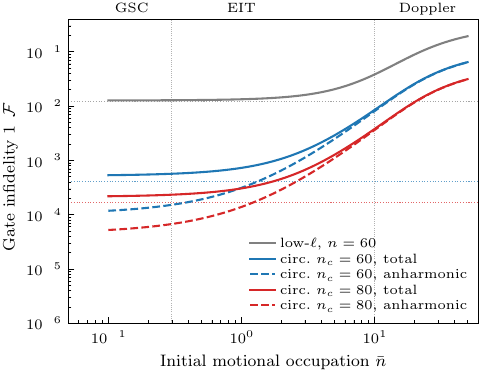}
\caption{\label{fig:circ_thermal}\textbf{Gate infidelity at the circular-Rydberg operating point.} Color encodes the Rydberg species/state: gray, low-$\ell$ $^3\!S_1$ at $n{=}60$ (main-text operating point, 4\,K cryogenic, $\tau_r\!=\!200\,\mu$s); green, circular $n_c{=}60$ ($\eta\!=\!1.04{\times}10^{-3}$); purple, circular $n_c{=}80$ ($\eta\!=\!6.9{\times}10^{-4}$). Line style encodes thermal/BBR environment and error decomposition: \emph{solid} = total with $300\,$K Purcell-adapted decay ($\tau{=}6\,$ms for $n_c{=}60$, $\tau{=}15\,$ms for $n_c{=}80$; based on Pultinevicius \emph{et al.}~\cite{Pultinevicius2025}); \emph{dash-dot} = total with $4\,$K cryogenic decay (radiative limit, $\tau\!=\!119\,$ms, $502\,$ms respectively); \emph{dashed} = anharmonic $C_4$ contribution only (decay removed, same for both environments). Dotted horizontal lines mark the corresponding Rydberg-decay floors for each $(n, T_\mathrm{env})$ combination. Under EIT-cooled operation ($\bar n\!\approx\!1$), circular $n_c{=}60$ sits at $1\!-\!\mathcal{F}\!\approx\!7\!\times\!10^{-4}$ at $300\,$K or $3\!\times\!10^{-4}$ at $4\,$K, both well below the low-$\ell$ $n{=}60$ baseline of $1.4\!\times\!10^{-2}$. The dashed curves (anharmonic alone) shift downward by the expected factor $(\eta_\mathrm{circ}/\eta_\mathrm{low\text{-}\ell})^2\!\approx\!0.30$ relative to the low-$\ell$ reference, directly verifying the $\eta^2$ scaling argument of Sec.~S7. Top-axis labels mark cooling regimes as in Fig.~\ref{fig:thermal}. The $4\,$K cryo curves merge with the dashed (anharmonic-only) curves over most of the plot, confirming that cryogenic operation renders the gate essentially anharmonic-limited.}
\end{figure}

\paragraph*{Local-circular architecture: circular only near the ion.}
The atom need not be in a circular state throughout the full protocol. Circular preparation is required only during the ${\sim}\!5\,\mu$s atom--ion CZ itself, i.e., when the atom is at the gate location $d_\mathrm{CZ}\!\approx\!12\,\mu$m from the ion. Everywhere else (tweezer loading, atom--atom Rydberg CZ operations inside the atom array, and ballistic shuttling of the tweezer between the array and the ion interface) the atom remains in either its ground state $^3P_0$ or a conventional low-$\ell$ Rydberg state. The electric-field sensitivity of the circular manifold is therefore confined to a ${\sim}\!5\,\mu$s window at the single gate location, where the local environment can be independently engineered (e.g., with a local RF-quiet zone in the trap geometry, or a dedicated Faraday aperture near the ion trap surface). This is simpler than maintaining circular purity across the full multi-hundred-micron atom array. In particular, the Purcell-suppressed BBR lifetime demonstrated in Ref.~\cite{Pultinevicius2025} is a property of the tweezer surrounding the atom at the gate location: the atom is only Yb-circular while it is inside that purpose-built tweezer, and the rest of the hybrid apparatus sees low-$\ell$ Rydberg or ground-state Yb with their standard field sensitivities.

\paragraph*{Sign-flipping the circular $C_4$ via microwave dressing.}
The multi-ion mode-closure scheme of Sec.~S5 requires a toggle between attractive ($+C_4$) and repulsive ($-C_4$) atom--ion interactions. For low-$\ell$ Rydberg this was achieved by fast microwave transfer between $^3\!S_1$ (attractive) and Stark-dressed $^3\!D$ (repulsive via Peper dressing~\cite{Peper2025}). Within the circular manifold the analogous mechanism is an off-resonant microwave drive of the $|n,c\rangle\!\leftrightarrow\!|n{+}1,c\rangle$ transition, which has been experimentally demonstrated as a means to tune and null Rydberg polarizability on the ${>}80\%$ level~\cite{Sillitoe2023}.

Let $\omega_d$ be the drive frequency, $\Omega_d$ its Rabi coupling to the $|n,c\rangle\!\to\!|n{+}1,c\rangle$ transition, and $\delta_d\!=\!\omega_d\!-\!\omega_{n,n+1}$ the detuning from resonance, where $\omega_{n,n+1}\!\equiv\!(E_{n+1,c}\!-\!E_{n,c})/\hbar\!\approx\!2\mathrm{Ry}/(\hbar n^3)$ is the circular-to-circular transition frequency (microwave band, ${\sim}\!30\,$GHz at $n{=}60$). In the dispersive regime $|\delta_d|\!\gg\!|\Omega_d|$, the drive Autler--Townes shifts both dressed states and modifies their dc polarizability by
\begin{equation}\label{eq:alpha_RF}
    \alpha_\mathrm{eff}(\delta_d) \;=\; \alpha_\mathrm{circ}^{(0)} \;+\; \frac{|\Omega_d|^2}{\delta_d}\,\frac{|\braket{n{+}1,c|\hat d|n,c}|^2}{\hbar^2\,\delta_d}\;-\;\ldots,
\end{equation}
where $\hat d$ is the electron electric-dipole operator and the second term carries the sign of $\delta_d$. Blue-detuning ($\delta_d\!>\!0$) adds a positive dressing contribution and enhances $\alpha_\mathrm{eff}$; red-detuning ($\delta_d\!<\!0$) subtracts from the bare polarizability. For $|\Omega_d|/2\pi\!\sim\!100\,$MHz and $|\delta_d|/2\pi\!\sim\!1\,$GHz, well inside the dispersive regime and within commercial microwave-source specifications at 30\,GHz, the dressing correction is $|\alpha_\mathrm{RF}|/\alpha_\mathrm{circ}^{(0)}\!\sim\!1$--$10$, sufficient to flip the sign of $\alpha_\mathrm{eff}$ by the choice of detuning. The resulting $C_4$ force reverses direction mid-gate, as required by the Rydberg-toggling protocol of Eq.~\eqref{eq:closure}.

\paragraph*{Toggling timescale and bandwidth.}
The limiting timescale for switching $C_4$ sign is set by: (i) the microwave-source amplitude modulation bandwidth (standard commercial $30\,$GHz synthesizers deliver ${\sim}\!1\,$ns rise/fall), and (ii) the dispersive-regime validity window $|\Omega_d|\!\ll\!|\delta_d|$, which requires the drive to be ramped adiabatically over at least $2\pi/|\delta_d|\!\sim\!1\,$ns to avoid populating $\ket{n{+}1,c}$. Both constraints give a ${\sim}\!1$--$2\,$ns minimum toggle time, identical to the microwave $\pi$-pulses used for ${}^3P_2\!\leftrightarrow\!{}^3D_2$ toggling in the main-text low-$\ell$ protocol. The multi-ion mode-closure schedules of Sec.~S5 (Table~\ref{tab:toggling}, up to $17$ toggles per gate for $N{=}10$ ions) therefore carry over unchanged to the circular operating point with ${\lesssim}\!340\,$ns of pulse overhead, negligible compared to the $5\,\mu$s gate duration. The required additional hardware at the gate site is one $\sim\!30\,$GHz microwave source with amplitude and detuning modulation, nominally identical to the microwave infrastructure already needed for circular-state preparation via adiabatic rapid passage~\cite{Nguyen2018,Cortinas2020}.

\paragraph*{Practical trade-offs of the circular upgrade.}
With the local-circular architecture, the remaining overheads are limited to the gate location itself:
\begin{enumerate}\itemsep0pt
\item \emph{Preparation time} ($10$--$100\,\mu$s): circularization proceeds through adiabatic microwave sweeps traversing the Stark manifold~\cite{Nguyen2018,Cortinas2020}. For the atom-shuttle protocol, this overhead is absorbed into the ${\sim}\!200\,\mu$s transit time to the ion and does not reduce the entanglement rate. Room-temperature tweezer-based circularization has been demonstrated~\cite{Pultinevicius2025}, eliminating the need for a cryogenic apparatus.
\item \emph{DC field stability at the gate location}: the circular state's larger polarizability ($\alpha\!\approx\!20\!\times\!\alpha_{^3\!S_1}$) translates to an enhanced second-order Stark shift of ${\sim}\!17\,$MHz at the ion's field at $d_\mathrm{CZ}\!=\!11.7\,\mu$m, compared to ${\sim}\!10\,$MHz for low-$\ell$. The fractional sensitivity to separation fluctuations $\delta V/\delta d\!=\!4V_0/d$ is essentially unchanged (6.0 vs 6.2\,MHz/$\mu$m), so the gate phase is no more sensitive to $d$-jitter than the main-text protocol.
\item \emph{Magnetic-field sensitivity}: the circular state's Zeeman shift scales as $\mu_B m_j \!\approx\!\mu_B(n{-}1)$, making it ${\sim}\!30{\times}$ more sensitive than $^3\!S_1$. Typical ion-trap apparatuses achieve $\mu$G-level field stability, keeping the associated dephasing below $10^{-4}$ over the $5\,\mu$s gate.
\item \emph{Multi-ion Rydberg-toggling}: the ${}^3P_2\!\leftrightarrow\!{}^3D_2$ toggling of Sec.~S5 carries over to the circular manifold via RF/microwave dressing of the circular-to-circular transition. An off-resonant microwave drive at $\omega_d$ slightly detuned from the $n\!\leftrightarrow\!n{+}1$ circular transition (${\sim}\!30\,$GHz at $n_c{=}60$) Autler-Townes dresses $|n,c\rangle$ with $|n{+}1,c\rangle$, adding a dressing contribution $\alpha_\text{RF}\propto |\Omega_d|^2/(\omega_d-\omega_{n,n+1})$ to the bare polarizability. The sign of $\alpha_\text{RF}$ changes with the sign of the detuning, so a modest dressing amplitude produces a dressed circular state with either sign of effective $C_4$, a structurally different mechanism from the bare-state-pair toggle of the low-$\ell$ protocol (Sec.~S6) that achieves the same microwave-driven sign reversal. This RF-dressing mechanism has been demonstrated with ${>}\,80\%$ polarizability modulation in neighboring-Rydberg experiments on Cs~\cite{Sillitoe2023}, and the sub-ns switching of the microwave-dressing envelope is fully compatible with the multi-segment toggle schedules of Sec.~S5. Thus the circular operating point retains multi-ion compatibility at the price of one additional ${\sim}\!30\,$GHz RF source per gate site.
\item \emph{Species specificity}: Ref.~\cite{Pultinevicius2025} demonstrates circular-state preparation and trapping at 300\,K in a single species, validating the technique; porting to \Yb requires characterization of the $6s{\cdot}nc$ ladder but no new principles.
\end{enumerate}
With these points taken into account, the circular \Yb upgrade is a drop-in replacement for the single-ion atom--ion CZ at the gate location, projecting $\mathcal{F}\!\approx\!99.85\%$ at room temperature without cryogenic shielding or extended RF isolation of the atom array. This is a simpler experimental path than the $^3\!S_1$ operating point when the single-gate fidelity itself matters (e.g., for protocols that do not rely on erasure-aware QEC).

\paragraph*{10-ion crystal gate with the circular upgrade.}
The multi-ion closure scheme of Sec.~S5 combined with the microwave-dressed $\sigma^+\!/\sigma^-$ circular toggling above extends the circular gate to ion crystals. Using the identical toggle schedule of Table~\ref{tab:toggling} (the closure residual $|\alpha_m(T)|^2\!\lesssim\!10^{-18}$ is independent of $\eta$ or $\tau_r$, depending only on the mode spectrum $\{\omega_m\}$), the 10-ion gate infidelity decomposes into three channels as for the single-ion case: (i) Rydberg decay (motion-independent, same per-gate value as Table~\ref{tab:circular}); (ii) anharmonic $C_4$ dephasing; (iii) technical errors. The multi-mode anharmonic scales as
\begin{equation}\label{eq:multimode_anharm}
    \mathrm{Anharm}_N \;=\; \left(\frac{10\pi\,\eta\,\omega_g\,\bar n}{\omega_\mathrm{COM}}\right)^{\!2}\!\sum_{m=1}^{N}\frac{b_k[m]^4}{\omega_m^2},
\end{equation}
where $\omega_\mathrm{COM}\!=\!\omega_1\!=\!\omega$ is the axial center-of-mass mode frequency, $b_k[m]$ is the participation amplitude of the addressed (target) edge ion with index $k$ in axial mode $m$ (obtained by diagonalizing the ion-chain Hessian, normalized as $\sum_k b_k[m]^2\!=\!1$), and the double power $b_k[m]^4$ arises because both the Rydberg-conditional frequency shift $\delta\omega_m^{(r)}\!\propto\!b_k^2/\omega_m$ and the summed-incoherently squared phase error pick up one factor of $b_k^2$ each. For the 10-ion linear crystal the participation-weighted sum is $\sum_m b_k[m]^4/\omega_m^2\!\approx\!0.10$, so the multi-mode anharmonic is suppressed by an order of magnitude compared to the single-mode reference value. For large $N$ with the edge ion addressed, $b_k[m]^2\!\sim\!2/N$ uniformly across modes and the sum scales as $\langle 1/\omega_m^2\rangle/N\!\propto\!1/N$, so the anharmonic contribution decreases as $1/N$. Table~\ref{tab:crystal_circular} tabulates the 10-ion gate infidelity across the four operating regimes of Table~\ref{tab:circular}, plus the low-$\ell$ paper baseline.

\begin{table}[h]
\caption{\label{tab:crystal_circular}Projected per-gate infidelity for the 10-ion crystal atom--ion CZ using circular Rydberg \Yb, computed from Eq.~\eqref{eq:multimode_anharm}. ``Mode closure'' is the toggle-schedule residual ($\lesssim\!10^{-18}$, suppressed in the total column). The low-$\ell$ row is the main-text single-ion baseline replicated to the 10-ion crystal. ``Tech'' standard matches Table~\ref{tab:errors} ($\sim\!10^{-3}$); ``opt.\ tech'' is state-of-the-art cryogenic-trap ($\sim\!3\!\times\!10^{-4}$). Crystal-level anharmonic is suppressed by $\sum_m b_k[m]^4/\omega_m^2\!\approx\!0.10$ relative to the single-mode value, reflecting the dilution of the Rydberg-conditional frequency shift across all $N$ modes via the $b_k^4$ participation weights.}
\begin{ruledtabular}
\begin{tabular}{lcccc}
Regime & Decay & Anharm.\ (10-mode) & Tech & $1-\mathcal{F}$ ($\mathcal{F}$) \\
\hline
Low-$\ell\ ^3\!S_1$, 4\,K, EIT ($\bar n{=}1$) & $1.2\!\times\!10^{-2}$ & $2.4\!\times\!10^{-5}$ & $10^{-3}$ & $1.3\!\times\!10^{-2}$ (98.66\%) \\
Circ.\ $n_c{=}60$, 300\,K Purcell, EIT        & $4.2\!\times\!10^{-4}$ & $7.2\!\times\!10^{-6}$ & $10^{-3}$ & $1.4\!\times\!10^{-3}$ ($\mathbf{99.86\%}$) \\
Circ.\ $n_c{=}60$, 4\,K cryo, EIT             & $2.1\!\times\!10^{-5}$ & $7.2\!\times\!10^{-6}$ & $10^{-3}$ & $1.03\!\times\!10^{-3}$ ($\mathbf{99.90\%}$) \\
Circ.\ $n_c{=}60$, 4\,K, GSC ($\bar n{=}0.1$), opt.\ tech & $2.1\!\times\!10^{-5}$ & $7.2\!\times\!10^{-8}$ & $3\!\times\!10^{-4}$ & $3.2\!\times\!10^{-4}$ ($\mathbf{99.97\%}$) \\
Circ.\ $n_c{=}80$, 4\,K, GSC, opt.\ tech      & $5.0\!\times\!10^{-6}$ & $3.2\!\times\!10^{-8}$ & $3\!\times\!10^{-4}$ & $3.1\!\times\!10^{-4}$ ($\mathbf{99.97\%}$) \\
\end{tabular}
\end{ruledtabular}
\end{table}

The three bold entries (room-$T$ Purcell, $4\,$K EIT, and $4\,$K GSC with optimized technical budget) bracket the practical operating space. For non-cryogenic apparatuses the circular upgrade already reaches $\mathcal{F}\!\approx\!99.85\%$ for a ten-ion chain, ${\sim}\!8\times$ better than the low-$\ell$ baseline with no additional cooling hardware. Adding a $4\,$K BBR enclosure (standard in modern Paul-trap laboratories) removes the BBR decay contribution and brings the gate to $\mathcal{F}\!\approx\!99.89\%$ at EIT cooling. A further push to Raman GSC and a technical-error budget ${<}\!3\!\times\!10^{-4}$ reaches $\mathcal{F}\!\approx\!99.97\%$, at which point the crystal gate is no worse than the single-ion version (the multi-mode anharmonic contribution is suppressed) and is limited by residual technical errors. The 10-ion atom--ion interconnect then operates well below the surface-code threshold, as required for scaled fault-tolerant hybrid architectures.

\end{document}